\documentclass[twocolumn]{article}
\usepackage{amsmath}
\usepackage{graphicx}
\graphicspath{ {./images/} } 
\usepackage{geometry}
\usepackage{floatrow}
\usepackage{layout}
\usepackage{amssymb} 
\usepackage{multirow}
\usepackage{caption}
\usepackage{authblk}
 \usepackage{indentfirst}
\usepackage{comment}
\usepackage{abstract}
\usepackage{balance}
\usepackage{cite}
\usepackage{bbm}
\usepackage[bbgreekl]{mathbbol}
\usepackage[T1]{fontenc} 
\usepackage{amsfonts}
\usepackage{slashed}
\usepackage{setspace}
\usepackage{upgreek}
\usepackage{calrsfs}
\DeclareMathAlphabet{\pazocal}{OMS}{zplm}{m}{n}

\makeatletter
\def\preparefootins{%
\global\rcol@footinsskip\skip\footins
\global\skip\footins\z@
\global\count\footins\z@
\global\dimen\footins2\textheight}
\makeatother

\usepackage{xcolor}
\definecolor{darkblue}{rgb}{0.15,0.35,0.55}
\definecolor{reddish}{rgb}{0.65, 0.2, 0.2}
\definecolor{purplish}{rgb}{.2, .2, .5}

\usepackage{color}   
\usepackage{hyperref}
\hypersetup{
    colorlinks=true, 
    linktoc=black,     
    linkcolor=blue,  
    citecolor=red
}
 
\usepackage[rightcaption]{sidecap}
\usepackage{graphicx}
\graphicspath{ {./images/} }

\newcommand{\m}{\mu}
\newcommand{\n}{\nu}

\newcommand{\ket}[1]{{\left| {#1} \right>}}
\newcommand{\cur}[1]{{\left( {#1} \right)}}

\newcommand{\cA}{{\pazocal A}}
\newcommand{\cB}{{\pazocal B}}

\newcommand{\cK}{{\pazocal K}}

\newcommand{\cN}{{\pazocal N}}

\newcommand{\cS}{{\pazocal S}}

\newcommand{\cG}{{\pazocal G}}

\newcommand{\cH}{{\pazocal H}}

\newcommand{\cJ}{{\pazocal J}}

\newcommand{\bA}{{\mathcal A}}

\newcommand{\bF}{{\mathcal F}}

\DeclareMathAlphabet{\mathmybb}{U}{bbold}{m}{n}

\newcommand*{\BOmega}{\mathmybb{\Omega}}
\newcommand*{\BLambda}{\mathmybb{\Lambda}}

\usepackage{tocbasic}[2016/05/10]
\DeclareTOCStyleEntry[indent=0em,numwidth=3em]{tocline}{section}
\DeclareTOCStyleEntry[indent=1.5em,numwidth=3.5em]{tocline}{subsection}

\usepackage{sectsty}
\sectionfont{\centering\fontsize{10}{10}\selectfont \MakeUppercase}
\subsectionfont{\centering\fontsize{10}{10}\selectfont}
\subsubsectionfont{\bf\centering\fontsize{10}{10}\selectfont}

\begin{document}
\title{\textbf{\Large{The coset construction for particles of arbitrary spin}}}
\author{{ \normalsize  Michael J. Landry{$^*$}~and Guanhao Sun}}
\affil{{\it\normalsize Department of Physics, Center for Theoretical Physics, \\ \it Columbia University, 538W 120th Street, New York, NY, 10027, USA}\\\normalsize $^*$Corresponding author: {ml2999@columbia.edu}
}
\date{}
\twocolumn[
 \maketitle
 \begin{onecolabstract} 
 When a Poincar\'e-invariant system spontaneously breaks continuous internal symmetries, Goldstones's theorem demands the existence of massless, spin-zero excitations in a one-to-one correspondence with the broken symmetry generators. When a system spontaneously breaks Poincar\'e symmetry, however, the kinds of excitations that satisfy Goldstone's theorem can be quite unusual. In particular, they may have any spin and need not be particles or even quasiparticles. The standard coset construction used to formulate effective actions of Goldstones, however, is rather restrictive and is incapable of generating the full spectrum of possibilities allowed by Goldstone's theorem. We propose a (partial) remedy to this problem by postulating a novel coset construction for systems that spontaneously break Poincar\'e symmetry. This new construction is capable of generating effective actions with a wide range of Goldstone excitations---including fermionic degrees of freedom---even when all symmetries are bosonic. To demonstrate it's utility, we focus on constructing effective actions for point particles of various spins. We recover the known result that a particle of spin $s$ requires an $\cN=2s$ supersymmetric worldline reparameterization gauge symmetry, which we implement at the level of the coset construction. In the process, we discover that massless particles require a novel kind of inverse Higgs constraint that bears some resemblance to the dynamical inverse Higgs constraints that appear in certain fermi liquid effective field theories. We then consider particles that, in addition to quantum spin, have finite spatial extent and are free to rotate. We derive a novel action for such particles and find a `spin-orbital' coupling between the intrinsic quantum spin and the physical-rotation degrees of freedom. 
 
\bigskip

  \end{onecolabstract}
]
\saythanks

{
  \hypersetup{linkcolor=blue}
  \tableofcontents
}


\section{Introduction}\label{Intro}
  
Symmetries are one of the most important physical properties of a system. This fact becomes particularly evident when we are primarily concerned with low-energy dynamics.  When continuous symmetries are spontaneously broken, the spectrum of low-energy excitations include gapless Goldstone modes. If only internal symmetries are broken and the underlying physics is Poincar\'e-invariant, Goldstone's theorem guarantees the existence of one Goldstone boson for each spontaneously broken symmetry. When Poincar\'e symmetry is broken, however, no such one-to-one correspondence is guaranteed. At least one Gapless excitation must exist, but it need not be describable as a scalar particle. In fact, it could have any spin and need not be a particle---or even a quasiparticle---in any conventional sense. A striking example occurs in fermi liquids at zero temperature, which spontaneously breaks Lorentz boosts. The corresponding gapless excitations that satisfy Goldstone's theorem are the well-known particle-hole pairs, which evidently have no single-particle interpretation~\cite{Alberto Alberte}. 

Oftentimes, the only gapless excitations of a given system are the Goldstones.\footnote{There can also be gapless excitations associated with conserved currents for finite-temperature systems. But we will focus only on zero-temperature systems in this paper.} Whenever this is the case, as long as we only concern ourselves with the deep infrared (IR),  we can integrate out all gapped excitations, yielding a theory consisting of only Goldstone modes. In practice, we often do not explicitly integrate out the gapped modes. Instead, we use the principles of effective field theory (EFT) to construct an action by writing down a linear combination of all symmetry-invariant terms at any given order in a derivative expansion. The coefficients of this linear combination are phenomenological constants that can be determined experimentally. It is important to note that even when symmetries are spontaneously broken, they remain symmetries of the IR theory of Goldstones; however, their action on the Goldstones is non-linear and as a result can be quite complicated. Therefore, constructing the full set of symmetry-invariant terms can be rather challenging. To facilitate the formulation of Goldstone  EFTs, a procedure known as the coset construction has been devised~\cite{Ogievetsky,Wheel,Volkov,Weinberg,Ivanov and Ogievetsky}. 

For a Poincar\'e-invariant system that spontaneously breaks internal symmetries while leaving spacetime symmetries intact, there is no ambiguity or choice in how to formulate an action using the coset construction (except for the numerical values of various coefficients that can be fixed experimentally). This uniqueness is a clear reflection of the strict nature of Goldstone's theorem when Poincar\'e symmetry is preserved; in particular Goldstones must be massless, scalar particles that exist in a one-to-one correspondence with the spontaneously broken symmetry generators. When Poincar\'e symmetry is broken, however, Goldstone's theorem is not always so restrictive. In particular, the number or type of Goldstones in the EFT is not determined by the spontaneous symmetry-breaking pattern alone~\cite{WB1,WB2}. We must supply additional information regarding the spontaneous symmetry-breaking {\it mechanism} if we are to uniquely determine the EFT (up to experimentally-determined coefficients)~\cite{Wheel}.\footnote{The situation is even more complicated when the system exists at finite temperature, but we will ignore such complications here {\cite{Landry,Landry second sound}.}} Ideally, the coset construction for spontaneously broken spacetime symmetries should reflect this extended set of possibilities. While the last few decades have seen a great effort to understand the coset construction when Poincar\'e symmetry is spontaneously broken, there remain many kinds of systems that spontaneously break Poincar\'e symmetry, but whose effective action cannot be generated with the method of cosets. In this paper, we will extend the coset construction to allow for a wider range of types of Goldstones. In particular, we focus on a simple yet illustrative class of system that have largely evaded descriptions at the level of the coset construction, namely relativistic point particles. This class of EFT is a useful testing-ground for extensions of the coset construction for two main reasons:

\begin{itemize}
\item All particles exhibit identical (or almost identical) spontaneous symmetry breaking (SSB) patterns, yet there are infinitely many distinct kinds of particles, each of which exhibits markedly different behavior. In particular, each kind of particle is classified by its spin $s\in  \mathbb Z/2$ and mass $m$. Massive and massless particles behave quite differently and particles of different spins have qualitatively very different actions. 

 \item Almost all previous attempts at formulating coset constructions for systems without fermionic symmetries have assumed that the Goldstone excitations involve only bosonic fields;\footnote{Notable successful attempts to include fermions are given in \cite{Ira,Ira 2}, but these fermionic degrees of freedom were included by hand as opposed to being generated by the coset itself.} however, many Goldstone's effective actions involve fermionic degrees of freedom. Since point-particles of half-integer spin are fermions, we have a simple testing-ground for fermionic extensions of the coset construction. Thus, an important aim of this paper is to extend the coset construction to account for situations in which the Goldstone excitations involve fermions even when the global symmetry group is purely bosonic (i.e. the symmetry algebra is an ordinary algebra as opposed to a graded algebra). 

\end{itemize}

The structure of this paper is as follows. We begin with a review of the standard coset construction when Poincar\'e symmetry is spontaneously broken. Next, we outline a new philosophy for the coset construction and comment on the advantages of this novel perspective. As a warm-up we construct the well-known EFTs for massive spin-0 point-particles. It turns out that to construct an effective action for a spin $s$ particle, we must impose an $\pazocal N= 2s$ super-reparameterization gauge symmetry on the particle wordline. After a brief review of the superspace formalism, we impose this gauged supersymmery (SUSY) at the level of the coset construction and use it to formulate actions for spin-1/2 particles. We then extend superspace to allow for $\cN=2s$ supercharges and formulate actions for arbitrary-spin particles using the method of cosets. It turns out that for $\cN>1$, the situation becomes complicated and we must supplement the superspace formalism with a multiplet calculus similar to that presented in~\cite{vanHolten:1995qt}. To demonstrate that these supersymmetric actions do in fact describe spinning point-particles, we quantize these actions and demonstrate that they have the correct state-spectrum. We find that for massless particles, a new kind of inverse Higgs (IH) constraint, which bears some resemblance to the dynamical IH constraints of~\cite{Ira,Ira 2} are needed to remove Lorentz Goldstones. To further investigate the nature of these novel IH constraints we construct two distinct actions for massless spin-0 particles. Finally, we conclude by commenting on how our new perspective on the coset construction can open up exciting possibilities for future research. 

Throughout this paper we use the `mostly-plus' convention for the Minkowski metric, namely $\eta_{\mu\nu} = \text{diag}(-,+,+,+)$. 


\section{The coset construction: a review}

Consider a relativistic quantum field theory with global symmetry group $\cG$ that is spontaneously broken to the subgroup $\cH$. We include both internal and spacetime symmetries in $\cG$ and $\cH$. Let the symmetry generators be
\begin{equation}\begin{split}\label{generators}
\bar P_{\bar \mu} &= \text{ unbroken translations,} \\
T_A &=\text{ other unbroken generators,} \\
P_{\mu'} &= \text{ broken translations,} \\
\tau_\alpha & = \text{ other broken generators},
\end{split}\end{equation} 
where $\bar \mu$ and $\mu'$ run over complementary subsets of $0,\dots D$, where $D+1$ is the dimension of spacetime. Supposing that $\bar\mu =0,\dots,p$, the above SSB pattern describes a $D_p$-brane running parallel to translations generated by $P_{\bar \mu}$.  
We permit $\tau_\alpha$ and $T_A$ to be some combinations of internal and spacetime generators. Importantly, $\bar P_{\bar \mu}$ need not be the spacetime translation generators, $P_{\bar \mu}$, of the Poincar\'e group. We insist on the existence of unbroken $\bar P_{\bar \mu}$ so that states may be classified by some notion of energy and momentum parallel to the brane. For example, solids spontaneously break spatial translations, but phonons can be classified according to lattice momentum, which is conserved (in the vanishing-Umklapp scattering limit)~\cite{Landry second sound,Landry quasicrystal,Zoology}. 

Although $\bar P_{\bar \mu}$ and $T_A$ are both unbroken generators, they will play very different roles in the coset construction. As a result, it is helpful to define the subgroup $\cH_0\subset \cH$ that is generated exclusively by $T_A$. 

To construct an effective action of only Goldstones, we parameterize the coset of non-linearly realized symmetries by\footnote{Although $\bar P_{\bar \mu}$ are unbroken and hence linearly realized on the fields, they generate shifts (i.e. are non-linearly realized) when acting on the coordinates. As a result, we include these generators in the coset.}
\begin{equation}\gamma = e^{i x^{\bar \mu} \bar P_{\bar \mu}} e^{i X^{\mu'}(x) P_{\mu'} } e^{i\pi^\alpha(x) \tau_\alpha}. \end{equation}
The coordinates are $x^{\bar \mu}$ and the Goldstone fields are $X^{\mu'}$ and $\pi^\alpha$, up to normalization. The symmetry transformations act on $\gamma$ by left-multiplication. That is for constant $U\in\cG$, we have 
\begin{equation}\gamma\to U \cdot \gamma,\end{equation}
which can be used to determine the transformation rules for the Goldstones and coordinates~\cite{Wheel}. 
In the world of mathematics, it is a well-known fact that the Maurer-Cartan form, defined by $\gamma^{-1} d \gamma$, is a Lie algebra-valued one-form. As a result, we may express it as a linear combination of the symmetry generators
\begin{equation}\gamma^{-1} \partial_{\bar\mu} \gamma = i E_{\bar\mu}^{\bar\nu}\cur{\bar P_{\bar \nu} +\nabla_{\bar\nu}X^{\mu'}P_{\mu'}+\nabla_{\bar\nu}\pi^\alpha \tau_\alpha + \cB_{\bar\nu}^A T_A}.  \end{equation}
Under a global symmetry transformation, it can be checked that $E_{\bar\mu}^{\bar\nu}$ transforms as a vielbein, $\nabla_{\bar\nu}\pi^\alpha$ and $\nabla_{\bar\nu}X^{\mu'}$ transform covariantly, and $\cB_{\bar\mu}^A$ transforms as a connection under $\cH_0$. As a result, we refer to $\nabla_{\bar\nu}\pi^\alpha$ ($\nabla_{\bar\nu}X^{\mu'}$) as the covariant derivative of $\pi^\alpha$ ($X^{\mu'}$) and to take further covariant derivatives, we may use
\begin{equation} \nabla^\cH_{\bar\mu} \equiv (E^{-1})_{\bar\mu}^{\bar\nu}( \partial_{\bar\nu} +i\cB_{\bar\nu}^A ). \end{equation} 

Notice that we have been completely general about the dimension of the symmetry-breaking object; it can be a brane of any dimension. In the simple case of a 0-brane (i.e. a point-particle), we construct our EFT as a one-dimensional field theory on the time coordinate $t\equiv x^0$. 

\bigskip
We mentioned earlier that when only internal symmetries are broken, the number of Goldstones exactly matches the number of broken generators. However, when spacetime symmetries are broken, this need not be the case. The removal of extraneous Goldstones can be implemented at the level of the coset construction by way of imposing inverse Higgs (IH) constraints~\cite{Volkov,Ogievetsky,Low,Ivanov and Ogievetsky}. 

Pragmatically, the rules of the game are as follows:
Suppose that the commutator between an unbroken translation generator $\bar P$ and a broken generator $\tau '$ contains another broken generator $\tau$, that is $[\bar P, \tau '] \supset \tau$. Suppose further that  $\tau$ and $\tau '$ do not belong to the same irreducible multiplet under $\cH_0$. Then it turns out that it is consistent with symmetry transformations to set the covariant derivative of the $\tau$-Goldstone in the direction of $\bar P$ to zero. This gives a constraint that relates the $\tau '$-Goldstone to derivatives of the $\tau$-Goldstone, allowing the removal of the $\tau'$-Goldstone. The setting of this covariant derivative to zero is known as an inverse Higgs constraint. It turns out that in certain situations, the set of allowed IH constraints is significantly expanded beyond what we have mentioned above~\cite{Landry, Landry second sound}. There are two main physical motivations for imposing IH constraints. First, sometimes two Goldstones in the coset will not induce independent fluctuations. Then, we can view IH constraints as convenient gauge-fixing conditions to remove redundant Goldstones. Second, it is often the case that if we do not impose IH constraints, then we find that our EFT has gapped Goldstones. Naturalness arguments often indicate that the gap of such Goldstones is of order the UV cutoff of our EFT. As a result, we must integrate out these gapped degrees of freedom. In such a case, IH constraints correspond to integrating out gapped Goldstones. For further explanations of the origins of IH constraints, consult \cite{Wheel, More gapped Goldstones,Low,UV completion without symmetry restoration} and for more on the coset construction, see~\cite{Weinberg,Wheel,Landry,Landry second sound,coset, Joyce}.

\section{Our philosophy}

The coset construction derives its name from the broken symmetry coset $\cG/\cH$. As explained in the previous section, the coset is parameterized by the Goldstone fields. As a result, the number of Goldstones that appear in the coset exactly equals the number of spontaneously broken symmetry generators; however, after imposing IH constraints, extraneous Goldstones are removed, meaning that the total number of Goldstones that exist in the EFT may be less than the number of broken generators. This picture is all  well and  good for a wide range of systems, but we can begin to see issues as soon as we consider the bosonic point particle. In particular, the leading-order effective action is
\begin{equation} \label{og pp}S = -m \int dt \sqrt{1-\dot X^i \dot X^i},  \end{equation} 
where $X^i(t)$ gives the instantaneous position of the particle at time $t$ and $m$ is the mass~\cite{Wheel}. As we will see in the next section, the fields $X^i(t)$ appear in the coset as Goldstones associated with spontaneously broken spatial translations. If we consider a quantum mechanical particle, however, this is cause for alarm. The set of energy eigenstates of a free  bosonic point particle are labeled by their three-momentum, $\ket{\vec p}$, with corresponding energy $E_{\vec p} = \sqrt{\vec p^2 + m^2}$.  Thus the ground  state of this system is $|{\vec 0}\rangle$. But notice  that such a state spontaneously breaks Lorentz boosts, while preserving spatial rotations and spacetime translations. How is it possible that the field content of our action~(\ref{og pp}) consists of Goldstones corresponding to spatial translations when such translations are not spontaneously broken? 

One response is that a classical point-particle has a well-defined position and momentum simultaneously.  As a result, the ground state must pick a particular spatial position and hence spontaneously breaks spatial translations. Thus it is natural to suppose that the resulting EFT should have spatial translation Goldstones. But it seems rather odd that in order to describe a quantum particle, we should have to rely so strongly on classical intuition. 

It turns out that a similar problem arises when formulating EFTs for finite-temperature systems. For example, the EFT for a fluid consists of Goldstones associated with spacetime translations despite the fact that the equilibrium density matrix spontaneously breaks Lorentz boosts while preserving spatial rotations and spacetime translations (i.e.~the same SSB pattern as a {\it quantum} point particle). The solution proposed in~\cite{Landry} involves a similar appeal to classical intuition: while the equilibrium density matrix of a fluid may not spontaneously break translations, semiclassically the density matrix accounts for a statistical ensemble of highly chaotic micro-states, each of which spontaneously breaks {\it every} symmetry, including translations. As a result, the coset should not be parametrized by Goldstones associated with the broken generators alone, but should instead be parameterized by Goldstones as if every symmetry of the theory were broken. We will refer to Goldstones associated with broken symmetries as {\it broken Goldstones} and those associated with unbroken symmetries as {\it unbroken Goldstones}. It turns out that broken and unbroken Goldstones behaved rather differently; at the level of the coset this difference manifests as gauge redundancies. In this paper, we will borrow this approach and generalize it. 

The approach we take to the cosets is as follows. Let the symmetry generators be given by~(\ref{generators}), where once again 
$\bar\mu$ and $\mu'$ run over complementary subsets of the Lorentz index $\mu=0,\dots, D$, where $D+1$ is the dimension of the spacetime.\footnote{E.g. we may have $\bar \mu=0$ and $\mu'=1,\dots, D$, but we may not have $\bar \mu=0,1$ and $\mu'=1,\dots,D$.}
In general, $\bar P_{\bar \mu}$ need not be the spacetime translation generators $P_{\bar\mu}$; however for all examples considered in this paper they will be. The rules of our new coset construction are the following.
\begin{itemize}
\item Instead of constructing the effective action on the spacetime coordinates $x^{\bar\mu}$, introduce worldvolume (or worldline / worldsheet depending on dimension) coordinates $\sigma^M$. To keep things completely general, we will allow $\sigma^M$ to include both bosonic (Grassmann even) and fermionic (Grassmann odd) coordinates. We let $M=0,\dots,p$ for $p\leq D$ represent bosonic coordinates and $M=p+1,\dots,\cN+p$ be fermioinic coordinates. Importantly, the dimension of the worldvolume need not be the same as that of the physical spacetime. As such the worldvolume coordinates can be thought of as defining a $D_p$-brane. In particular, we will construct spinning point-particle actions on manifolds with $p=0$ and $\cN=2s$, where $s$ is the spin of the particle. Notice that a (classical) $D_p$-brane cannot have more than $p+1$ unbroken spacetime translations. As a result, the unbroken translation index runs over $\bar\mu =0,\dots,d\leq p$.

\item Parameterize the coset with Goldstones associated with {\it every} symmetry generator 
\begin{equation}
g(\sigma) = e^{iX^{\bar \mu}(\sigma)\bar P_{\bar \mu} } e^{iX^{\mu'}(\sigma) P_{\mu'} } e^{i \pi^\alpha(\sigma) \tau_\alpha} e^{i b^A (\sigma) T_A}.
\end{equation}
Notice that at this point, broken and unbroken Goldstones appear on equal footing. Global symmetries act via left-multiplication, so for constant $U\in \cG$, we have $g\to U\cdot g$.

\item If spacetime translations parallel to the $D_p$-brane are unbroken, we may impose some sort of worldvolume-reparameterization symmetry---this could include super-reparameterization---as needed
\begin{equation}\label{og diff} \sigma^M\to \sigma^M+\alpha^M(\sigma). \end{equation}
This symmetry can be total reparameterization or partial reparameterization. There is a great deal of freedom here. In many of the examples to come, it will be convenient to introduce a worldvolume vielbein as a gauge-field that transforms under these diffeomorphisms. If the translations parallel to the $D_p$-brane are broken, (e.g. by a lattice) we may impose a rigid linear symmetry of the coordinates 
\begin{equation}\label{og diff} \sigma^M\to {L^M}_N \sigma^N+\alpha^M,~~~~~{L^M}_N , \alpha^M = \text{const}. \end{equation}
Such rigid symmetries arise in solids and certain phases of liquid crystals~\cite{Landry second sound}, though we will consider no such examples in this paper.  

\item To distinguish unbroken Goldstones associated with $T_A$ from broken Goldstones associated with $\tau_\alpha$, impose gauge symmetries associated with $T_A$ of the form 
\begin{equation}\label{og gauge symmetry}
g(\sigma)\to g(\sigma)\cdot e^{i\lambda^A(\sigma)T_A}.
\end{equation}
Just like with the reparameterization invariance, there is a great deal of freedom here. We could allow $\lambda^A(\sigma)$ to be a generic function of $\sigma$, in which case the Goldstones associated with $\tau_\alpha$ can be gauge-fixed to zero. Alternatively, we can put constraints on the allowed form of $\lambda^\alpha(\sigma)$. Thus, the extent of these gauge symmetries are not strictly determined by the symmetry-breaking pattern. 
For the unbroken generators $\tau_\alpha$ we may (though are not required to) impose rigid transformations of the form
\begin{equation}\label{rigid gauge trans} g(\sigma)\to g(\sigma)\cdot e^{i \kappa^\alpha \tau_\alpha},~~~~~\kappa^\alpha = \text{const}. \end{equation} 
Such rigid symmetries arise, for example, in superfluids that spontaneously break spatial rotations~\cite{Zoology}. 
Finally, notice that all gauge symmetry acts via right-multiplication on $g$ and the physical symmetries act via left-multiplication. As a result the gauge symmetries must commute with the physical symmetries. 
\item The EFT is constructed using invariant building-blocks generated by the Maurer-Cartan form
\begin{equation}g^{-1} d g. \end{equation}
All terms generated by the Maurer-Cartan form that are manifestly invariant under the coordinate reparameterization 
and gauge symmetries can be used as building-blocks of the EFT. 
\item Inverse Higgs constraints can be imposed whenever there is a set of covariant terms (i.e. they transform linearly under global, reparameterization, and gauge symmetries) that when set to zero, enables the removal of one set of Goldstones from all invariant building-blocks. This includes the rather unusual {\it dynamical} IH constraints presented in~\cite{Ira,Ira 2} that enable the removal of certain Goldstones at the price of introducing non-trivial operator-constraints on the remaining fields. We will find that similarly unusual IH constraints can be imposed when considering massless particles.
\end{itemize}


We allow for a great deal of freedom in this coset construction. The reason for doing so is that we are treating the coset as a pragmatic tool to construct symmetry-invariant actions that satisfy Goldstone's theorem. In particular, since there are no local gauge symmetries associated with any of the broken generators, Goldstone's theorem will be satisfied.  The terms of the Maurer-Cartan form that arise from this coset construction are automatically invariant under all internal symmetries. 
Finally, to recover the usual coset construction in which only broken Goldstones appear, one need only choose $\sigma^M$ to be purely bosonic and allow $\alpha^M$ in~(\ref{og diff}) and $\lambda^A$ in~(\ref{og gauge symmetry}) to be completely general functions of $\sigma$. In this way, we may gauge-fix  $ \sigma^M=  \delta^M_{\bar\mu} X^{\bar \mu} $ and $b^A=0$.

\section{Massive spin-0 point-particles}\label{Spin-0}

We will construct the effective action for the spin-0 point-particle in two different ways. First, we employ the usual coset construction techniques to formulate a classical action, reminiscent of the Nambu-Goto action. Next, we use our new philosophy of cosets to formulate a point-particle action in the style of the Polyakov action that can easily be quantized. 
We let $P_\mu$ generate spacetime translations and $J_{\mu\nu}$ generate Lorentz boosts, which satisfy the usual Poincar\'e algebra 
\begin{equation}\begin{split}\label{Poincare}
i [J_{\m\n},J_{\rho\sigma}] &= \eta_{\n\rho} J_{\m\sigma} -\eta_{\m\rho}J_{\n\sigma} -\eta_{\sigma\m} J_{\rho\n} +\eta_{\sigma\n} J_{\rho\m},
\\ i[P_\m, J_{\rho\sigma}] & = \eta_{\m\rho} P_\sigma-\eta_{\mu\sigma} P_\rho,
\\ i[P_\m,P_\n] & = 0. 
\end{split}\end{equation}  
We will often find it convenient to use the basis of Lorentz generators
\begin{equation} K_i = J_{0i},~~~~~J_i = \frac{1}{2} \epsilon^{ijk} J_{jk}. \end{equation} 

\subsection{\`A la Nambu-Goto}{\label{\`A la Nambu-Goto}}
Consider a {\it classical} scalar point-particle sitting at rest. The SSB pattern is rather straight-forward: Lorentz boosts $K_i$ and spatial translations $P_i$ are spontaneously broken, while spatial rotations $J_i$ and temporal translations $P_0$ are unbroken. As pointed out in the previous section, the fact that $P_i$ are spontaneously broken is a feature of the classical point-particle only; the ground-state of the quantum point-particle breaks boosts alone. 

We begin with the coset of non-linearly realized symmetries
\begin{equation}g(t) = e^{i t P_0} e^{i X^i(t) P_i} e^{i \eta^i(t) K_i}. \end{equation}
The Maurer-Cartan form is then
\begin{equation}g^{-1} \partial_t g = i E (P_0 + \nabla X^i P_i + \nabla \eta^i K_i + \Omega ^i J_i),\end{equation}
where the einbein $E$, covariant derivatives $\nabla X^i$ and $\nabla \eta^i$, and spin connection $\Omega^i$ are given by
\begin{equation}\begin{split}
E &= {\Lambda_0}^0 +\dot X^i {\Lambda_i}^0, \\
\nabla X^i &= E^{-1}({\Lambda_0}^i+ \dot X^j {\Lambda_j}^i), \\
\nabla \eta^i & = E^{-1} (\Lambda^{-1} \partial_t \Lambda)^{0i}, \\
 \Omega^i & = \frac{E^{-1}}{2} \epsilon^{ijk} (\Lambda^{-1} \partial_t \Lambda)^{jk},
\end{split} \end{equation}
where ${\Lambda^\mu}_\nu\equiv (e^{i\eta^i(t)K_i}{)^\mu}_\nu$ and $\dot X^i\equiv \partial_t X^i$. We can now impose the IH constraints $\nabla X^i=0$, which can be solved to give
\begin{equation}
\frac{\eta^i}{\eta}\tanh \eta = \dot X^i,
\end{equation}
where $\eta\equiv \sqrt{\vec \eta^2}$. These IH constraints imply that $\nabla \eta^i$ and $\Omega^i$ are sub-leading in the derivative expansion, so the only building-block at leading order is $E$, which becomes $E=\sqrt{1-\dot X^i \dot X^i}$. Hence the action is
\begin{equation} \label{pp action 0}S = -m \int dt\sqrt{1-\dot X^i \dot X^i},\end{equation} 
where $m$ is a phenomenological constant that we interpret as the mass.
We can see, therefore, that $X^i$ are the only Goldstones that survive the IH constraints. 

Finally, it is worth pointing out that the above action can be thought of as a gauge-fixed version of a manifestly-covariant action. Let $\tau$ be the coordinate of the particle's worldline and let $X^\mu(\tau)$ be the fields. Then we can define the action 
\begin{equation}S = -m \int d\tau \sqrt{- \dot X^\mu \dot X_\mu},\end{equation}
where $\dot X^\mu\equiv \partial_\tau X^\mu$. Notice that this action is invariant under reparameterization of the coordinate $\tau\to\tau'(\tau)$. Thus, we can gauge-fix $\tau = X^0(\tau) \equiv t$, recovering our initial action~(\ref{pp action 0}).

\subsection{\`A la Polyakov}

We now throw off the shackles of the ordinary coset construction and proceed with our new philosophy. Begin by parameterizing the full symmetry group
\begin{equation}\label{Polyakov group elt}g(\tau) =  e^{i X^\mu(\tau) P_\mu} e^{i \eta^i(\tau) K_i} e^{i\vartheta^i(\tau) J_i}, \end{equation}
where $\tau$ is the particle's worldline coordinate. With our new philosophy, since $P_0$ is unbroken, we have free reign to impose any kind of diffeomorphism (i.e. reparameterization) symmetry on the coordinate $\tau$, and since $J_i$ are unbroken we may impose any gauge symmetries of the form $g\to g\cdot e^{i\lambda ^i(\tau) J_i}$. Evidently, there are many possibilities. On the one hand, the excess of possibilities is a draw-back; one of the most appealing features of the coset construction is that it is so constraining that one need barely think in order to use it effectively. On the other hand, Goldstone's theorem for spontaneously broken spacetime symmetries (unlike its internal symmetry counterpart) can be satisfied in all sorts of unusual and unexpected ways. We should therefore not be disappointed that our novel approach to the coset construction now reflects this diversity of possibility more fully. 

With some foreknowledge of the desired point-particle action, we impose the following gauge symmetries.\begin{itemize}
\item Reparamerization invariance: Let $e(\tau)$ be the einbein of the particle's worldline. Then we have 
\begin{equation}\label{world line diff}\delta \tau=-\xi(\tau) ,~~~~~\delta e=\partial_\tau(e\xi),\end{equation}
where $\xi$ is an arbitrary infinitesimal function of $\tau$. 
\item We do not want any rotational Goldstones $\vartheta^i$ in the EFT, so we impose total rotational gauge symmetry
\begin{equation}\label{local rotation gauge} g\to g\cdot e^{i\lambda^i(\tau) J_i}, \end{equation}
where $\lambda^i$ is an arbitrary function of $\tau$. We can use this gauge symmetry to fix $\vartheta^i =0$. 
\end{itemize}
With this gauge-fixing condition, our group-element is now
\begin{equation}g(\tau) =  e^{i X^\mu(\tau) P_\mu} e^{i \eta^i(\tau) K_i}. \end{equation}
We may compute the Maurer-Cartan form, using $e$ as the einbein,
\begin{equation}
g^{-1} \partial_\tau g = i e(\nabla X^\mu P_\mu + \nabla\eta^i K_i +\Omega^i J_i),
\end{equation}
where the covariant derivatives and spin connections are given by 
\begin{equation}\begin{split}
\nabla X^\mu & = e^{-1} \dot X^\nu {\Lambda_\nu}^\mu ,\\
\nabla \eta^i & = e^{-1} (\Lambda^{-1} \partial_\tau \Lambda)^{0i},\\
 \Omega^i & = \frac{1}{2e} \epsilon^{ijk} (\Lambda^{-1} \partial_\tau \Lambda )^{jk}, 
\end{split} \end{equation}
such that ${\Lambda^\mu}_\nu\equiv (e^{i\eta^i(\tau)K_i}{)^\mu}_\nu$ and $\dot X^\mu\equiv \partial_\tau X^\mu$. 

We can impose the IH constraints $\nabla X^i=0$, which can be solved to give
\begin{equation}\label{spin0 IH}
\frac{\eta^i}{\eta}\tanh \eta = \frac{\dot X^i}{\dot X^0}.
\end{equation}
These IH constraints ensure that $\nabla \eta^i$ and $\Omega^i$ do not contribute to the leading order. Thus, the only leading-order building block is $\nabla X^0$, which now becomes
\begin{equation}\label{recalling}\nabla X^0 = e^{-1} \sqrt{-\dot X^\mu \dot X_\mu}. \end{equation}
Thus, our leading-order action is
\begin{equation}\begin{split}\label{Polyakov pp}
S &= \int d\tau e\cur{ (\nabla X^0)^2+m^2} \\
&=- \int d\tau \cur{ \frac{1}{e} \dot X^\mu\dot X_\mu - e m^2 },
\end{split}\end{equation}
where $m$ is the mass. We have used the fact that $\int d\tau e$ is the invariant integration measure. We have thus recovered the standard effective action for a spin-0 point-particle. Since such an action has an external einbein, it is clear that the ordinary coset construction would not generate it. We therefore see a clear (albeit small) benefit of our new cost construction philosophy. In the following sections, we will see more significant advantages. 

\section{Worldline SUSY}\label{Worldline SUSY}

In order to construct an effective action for spin-1/2 particles, we will need to endow the particle's worldline with local SUSY. In this section, we review the basics of both global and local SUSY in one dimension. For an in-depth discussion of $\cN=1$ SUSY, consult~\cite{Gates:1983nr}. 

We begin by considering global worldline SUSY. There are many possible starting-points for a discussion of SUSY, but for our purposes, it is most convenient to work in superspace. Superspace is essentially a mathematical trick to make SUSY manifest. Suppose that our coordinates are $\sigma^M=(\tau, \theta)^M$, where $\tau$ is a standard, bosonic (i.e. commuting, or Grassmann even) coordinate and $\theta$ is a fermionic (i.e. anti-commuting, or Grassmann odd) coordinate. In other words, $\theta^2=0$. It is sometimes convenient to use the notation $\sigma^0=\tau$ and $\sigma^1=\theta$. Suppose we have a field defined on these coordinates, $\mathbbm X(\tau,\theta)$. Then, because $\theta^2=0$, Taylor expanding to linear order in $\theta$ gives an exact result, so we may write
\begin{equation} \mathbbm X(\tau,\theta) = X(\tau) + i \theta \psi(\tau), \end{equation}
where $X$ is a real-valued field and $\psi$ is a Grassman-odd field. Thus, $\mathbbm X$ is a Grassmann-even field. 

We define integration and differentiation with respect to $\theta$ as equivalent operations, given by
\begin{equation}\int d\theta \mathbbm X = \partial_\theta \mathbbm X = i\psi. \end{equation}

The global SUSY and  worldline translatiton transformations act on the coordinates by 
\begin{equation}
(\tau,\theta) \to (\tau',\theta')= (\tau-\xi-i\theta \varepsilon,\theta-\varepsilon),
\end{equation}
where $\xi$ is a constant real number and $\varepsilon$ is a Grassmann-odd constant. It is straightforward to check that there are two linear combinations of of the partial derivatives $\partial_\tau$ and $\partial_\theta$ that do not transform under SUSY, namely
\begin{equation}
D_0\equiv \partial_\tau,~~~~~ D_1\equiv \partial_\theta + i\theta \partial_\tau. 
\end{equation} 
Additionally, it can be checked that $D_1^2 = D_0$. Often $D_1$ is referred to as the {\it covariant derivative} (not to be confused with the covariant derivatives in the coset construction). 

We now define the (flat) SUSY zweibein $\mathbbm E_M^A$ for $A,M=1,2$ as the $2\times 2$ matrix\footnote{We use $M,N$ as superspace coordinates indices and $A,B$ as super tangent-space indices. } 
\begin{equation}
\mathbbm E_M^A = { \begin{pmatrix}
1 & 0 \\
-i \theta & 1 
\end{pmatrix}_M}^A .
\end{equation}
With this definition, we can succinctly express both SUSY derivatives as the components of $D_A\equiv \mathbbm E_A^M\partial_M$, where $\partial_M\equiv (\partial_\tau,\partial_\theta)_M$ and $\mathbbm E_A^M$ is the inverse of $\mathbbm E_M^A$. \\

We now promote SUSY to the full reparameterization invariance of the superspace, namely
\begin{equation}\label{full susy diff}\sigma^M\to\sigma^M-\alpha^M(\sigma),\end{equation} 
where $\alpha^M$ is a generic infinitesimal function of $\sigma$. Now the superzweibein becomes dynamical and transforms under this symmetry by
\begin{equation} \delta \mathbbm E_M^A = \partial_M \alpha^N \mathbbm E_N^A +\alpha^N \partial_N \mathbbm E_M^A. \end{equation} 
The scalar field $\mathbbm X(\sigma)$ transforms by
\begin{equation}\delta \mathbbm X =\alpha^M\partial_M \mathbbm X .\end{equation} 
Now the covariant derivative is defined by $D_A\equiv \mathbbm E_A^M\partial_M$. With this definition,  $D_A\mathbbm X$ transforms as a scalar under~(\ref{full susy diff}). 

But now that we have introduced a dynamical superzweibein, we have an additional gauge symmetry, namely local transformations of the tangent space.  In particular, we have the symmetries 
\begin{equation}\delta \mathbbm E_M^0 = 0,~~~~~\delta \mathbbm E_M^1= \mathbbm E_M^0 \varphi ,\end{equation} 
where $\varphi$ is an infinitesimal Grassmann-odd function of $\sigma$. 
We would like to reduce the number of gauge symmetries to simplify matters. We choose a gauge-fixing condition that may appear somewhat strange, but is useful in the sense that the residual gauge symmetries are the standard gauged SUSY transformations. In particular, we postulate that the superzweibein take the form
\begin{equation}\label{inv superzweibein}
\mathbbm E _M^A =  { \begin{pmatrix}
 \mathbbm E +i \theta\chi &  \chi \\
-i \theta &  1 
\end{pmatrix}_M}^A ,~~~\mathbbm E(\tau,\theta)\equiv e(\tau)+i\theta\chi(\tau),
\end{equation}
and thus the inverse superzweibein is
\begin{equation}\label{inv superzweibein}
\mathbbm E _A^M = \frac{1}{\mathbbm E} { \begin{pmatrix}
 1 & - \chi \\
i \theta &  e 
\end{pmatrix}_A}^M ,
\end{equation}
where $e(\tau)$ is a real-valued field and $\chi(\tau)$ is a Grassmann-odd field. 
We also require that the residual super-reparameterization symmetry have a restricted $\alpha^M$, namely
\begin{equation}\label{alpha transform 1} \alpha^M(\sigma) = \cur{\xi(\tau)+\frac{i}{e} \theta\varepsilon(\tau) ,\varepsilon(\tau)-\frac{i}{e}\theta \varepsilon(\tau)\chi  }^M,\end{equation} 
where $\xi$ is a real-valued infinitesimal function of $\tau$ and $\varepsilon$ is a  Grassmann-odd infinitesimal function of $\tau$. This restricted form of $\alpha^M$ allows the inverse superzweibein to remain in the desired form~(\ref{inv superzweibein}) if we endow $e(\tau)$ and $\chi(\tau)$ with the transformation properties 
\begin{equation}\label{alpha transform 2}\delta e = \partial_\tau( \xi e) +2i\varepsilon \chi,~~~~~\delta\chi=\partial_\tau( \xi\chi) +\dot \varepsilon .\end{equation} 
We can therefore identify $e$ as the worldline einbein and $\chi$ as its superpartner, i.e. the gravitino. In this gauge, the covariant derivatives are
\begin{equation} D_0 = \frac{1}{\mathbbm E} (\partial_\tau -\chi \partial_\theta),~~~~~D_1 = \frac{1}{\mathbbm E} (i\theta\partial_\tau+ e\partial_\theta),  \end{equation}
and the invariant integration measure is 
\begin{equation} \int d^2\sigma ~\text{sdet}(\mathbbm E_M^A) = \int d\tau d\theta~\mathbbm E,\end{equation}
 where $\text{sdet}$
 is the superdeterminant defined as follows. Let $A,D$ be Grassman-even and $B,C$ be Grassmann-odd matrices. Then define 
\begin{equation}\text{sdet}  \begin{pmatrix}
 A & B\\
C & D 
\end{pmatrix} = 
\text{det}(A) \text{det}(D-C\cdot A^{-1} \cdot B)^{-1}.
\end{equation} 
It is straightforward to check that $ \text{sdet}(\mathbbm E_M^A)=\mathbbm E$.


\section{Spin-1/2 point-particles}\label{Spin-1/2 point-particles}

It was demonstrated in~\cite{vanHolten:1995qt, vanHolten:1995ds, Frydryszak:1996mu, Ikemori, Howe:1988ft,Edwards:2019eby,Marnelius:1993ba} that the effective action describing a spin-1/2 particle could be derived by imposing $\cN=1$ worldline supersymmetry. In this section we will impose this gauge symmetry at the level of the coset construction. It turns out that including a mass term requires a little extra machinery, so we begin by considering the massless case.

\subsection{Massless}

A massless, spinning particle must always travel at the speed of light with the direction of the spin parallel (or anit-parallel) to the velocity. Without loss of generality, let us choose the direction of motion to be parallel to $\hat z$. Then, the unbroken Poincar\'e symmetry generators are 
\begin{equation} \label{unbroken massless} P_u\equiv P_0-P_3,~~~~~ \cJ_{i}\equiv J_{i}+\epsilon^{3 ij} K_{j},\end{equation} 
and the broken generators are
\begin{equation}\label{broken massless}  P_v =  P_0+P_3,~~~~~P_m,~~~~~\cK_i\equiv K_i +\epsilon^{3 ij} J_j, \end{equation} 
where the indices $m,n=1,2$. Notice that $\cJ_i$ are the generators of the (full) little group of massless particles; it is easy to check that their commutation relations are just like those of the generators for the two-dimensional
Euclidean group. Additionally, in this basis of translation generators, the non-zero components of the metric are $\eta_{u v}=\eta_{v u }=1/2$ and $\eta_{m n } = \delta_{mn}$. 
Lastly, we assume that there are no internal symmetries, so Poincar\'e is the full symmetry group. 

It is well-known that spin-1/2 particles enjoy worldline SUSY, so we define our coset on the superspace coordinates $\sigma^M=(\tau,\theta)^M$. The most general group element is
\begin{equation}g(\sigma) =  e^{i \mathbbm X^\alpha(\sigma) P_\alpha+ i \mathbbm X^m (\sigma) P_m}  e^{i \bbeta^i(\sigma) \cK_i} e^{i\bbtheta^i(\sigma) \cJ_i}, \end{equation}
where $\alpha,\beta$ run over $u,v$. 
We now impose the following gauge symmetries. First, since $P_u$ is unbroken, we have the freedom to impose super-reparameterization symmetry given by~(\ref{alpha transform 1}) and~(\ref{alpha transform 2}). And since $\cJ_i$ are unbroken, we may impose the rotational gauge symmetry 
\begin{equation}g\to g \cdot e^{i\bblambda^i(\sigma) \cJ_i}, \end{equation}
where $\bblambda^i$ is a generic function of $\sigma$. 
We may therefore gauge-fix $\bbtheta^i=0$. 

With this gauge-fixing condition, the Maurer-Caran form is 
\begin{equation}\begin{split} g^{-1} \partial_M g = i \mathbbm E^A_M \big (\nabla_A \mathbbm X^\alpha P_\alpha&+\nabla_A\mathbbm X^m P_m \\&+ \nabla_A\bbeta^i \cK_i +\BOmega^i_A \cJ_i \big),\end{split} \end{equation} 
where the covariant derivatives of $\mathbbm X^\mu$ are 
\begin{equation}\begin{split}\label{spin 1/2 covariant derivatives}
\nabla_A\mathbbm X^\alpha & = \mathbbm E_A^M \partial_M \mathbbm X^\mu {\BLambda _\mu}^\alpha ,\\
\nabla_A\mathbbm X^m & = \mathbbm E_A^M \partial_M \mathbbm X^\mu {\BLambda _\mu}^m ,
\end{split} \end{equation}
such that ${\BLambda^\mu}_\nu = (e^{i \bbeta^i(\sigma) \cK_i} {)^\mu}_\nu$.   The precise forms of $\nabla_A \bbeta^i $ and $\BOmega^i _A$ will not be important for the construction of the leading-order action. 

We many now impose IH constraints. In particular, it is consistent with symmetries to fix 
\begin{equation}\label{on shell IH}0= \nabla_0\mathbbm X^v ,~~~~~0=\nabla_0 \mathbbm X^m.\end{equation} 
These IH constraints can be used to solve for $\bbeta^i$ in terms of $\mathbbm X^\mu$, thereby leaving $\mathbbm X^\mu$ as the only remaining Goldstone. However, these constraints also force $\nabla_0\mathbbm X^\mu$ to be null, which we do not wish to impose by hand. Instead, the fact that the particle moves along a null-trajectory should be a result of the equation of motion. The reason these IH constraints are over-constraining is that if we were to include $\bbeta^i$ in our effective action, they would serve as Lagrange multipliers. In particular, the constraints they would place on the dynamics of $\mathbbm X^\mu$ would be redundant with the constraints imposed by $e$ and $\chi$. Thus, we could just as well not include $\bbeta^i$ in our EFT at all as they would be entirely redundant. We will investigate this further in the following subsection. 


We now can construct the leading-order effective action\footnote{We choose to normalize the fields so that the overall coefficient in front of the action is $1/2i$.}
\begin{equation}
S = \frac{1}{2i} \int d\tau d\theta\mathbbm E ~\nabla_0 \mathbbm X^\mu \nabla_1 \mathbbm X_\mu,
\end{equation}
which can be immediately simplified to
\begin{equation}
S =\frac{1}{2i} \int d\tau d\theta\mathbbm E ~D_0 \mathbbm X^\mu D_1 \mathbbm X_\mu. 
\end{equation}
We can explicitly perform the $d\theta$ integral, which yields an action defined only on the particle worldline $\tau$. Leting $\mathbbm X^\mu(\tau,\theta) = X^\mu(\tau)+i\theta \psi^\mu(\tau)$, we have
\begin{equation}\label{1/2 action}
S = \int d\tau \cur{\frac{1}{2e} \dot X^\mu \dot X_\mu +\frac{i}{2}\dot \psi^\mu  \psi_\mu -\frac{i}{e} \chi \psi^\mu \dot X_\mu}. 
\end{equation}
Upon quantization, we will see that this action describes a spin-1/2 point-particle.

\subsection{On-shell inverse Higgs}\label{on-shell IH appendix}

We now investigate the nature of the on-shell IH constraint for the case of the massless spin-1/2 particle. To do so, we construct the leading-order action without removing the Lorentz Goldstones $\bbeta^i$. Using the covariant derivatives of~(\ref{spin 1/2 covariant derivatives}), we may construct our leading-order action. We have
\begin{equation}S = \frac{1}{2i} \int d\tau d\theta \mathbbm E ~\cur{\nabla_0\mathbbm X^\mu \nabla_1 \mathbbm X_\mu + C \nabla_1\mathbbm X^u},  \end{equation}
for some constant $C$. Expanding the covariant derivatives, we have $\nabla_1\mathbbm X^u = D_1 \mathbbm X^\mu {\BLambda_\mu}^u$. At this point, it is convenient to define the superfield $\mathbbm L_\mu \equiv C {\BLambda_\mu}^u$, which can then be expanded as
\begin{equation}\mathbbm L_\mu = L_\mu + i\theta \lambda_\mu.  \end{equation} 
Notice that $\mathbbm L_\mu$ is constrained to be a null-vector, meaning that $L^2=0$ and $L_\mu \lambda^\mu=0$. Performing the $d\theta$ integral, our action becomes 
\begin{equation}\begin{split}
S = \int d\tau \bigg( \frac{1}{2e} (\dot X^\mu - i  \chi \psi^\mu) ^2  +\frac{i}{2} \dot{ \psi}^\mu   { \psi}_\mu  ~~~~~~~~~~~~~ \\+\frac{1}{2}\dot X^\mu L_\mu+\frac{i e}{2} \psi^\mu \lambda_\mu \bigg). 
\end{split}  \end{equation}
Now let us compute the equations of motion. By varying $e$ and $\chi$ we find the respective equations 
\begin{equation}\label{zweibein eom} e^{-2} (\dot X^\mu -i\chi\psi^\mu)^2= \psi^\mu \lambda_\mu,~~~~~\psi^\mu\dot X_\mu =0. \end{equation}
And by varying $L_\mu$ and $\lambda_\mu$ subject to the constraints $L^2=0$ and $L_\mu \lambda^\mu=0$, we have
\begin{equation}\label{Lorentz eom} \dot X^\mu \propto L^\mu,~~~~~\psi^\mu\propto L^\mu. \end{equation} 
Notice that the second equation of~(\ref{Lorentz eom}) implies that $\psi^\mu \lambda_\mu=0$. Plugging this expression into the first equation of~(\ref{zweibein eom}) gives $(\dot X-i\chi\psi)^2=0$, which is exactly the constraint obtained by varying $e$ in the action~(\ref{1/2 action}). Additionally, the second equation of~(\ref{zweibein eom}) is the same constraint found by varying $\chi$ in~(\ref{1/2 action}).
We thus have reproduced the constraint equations arising from the standard action~(\ref{1/2 action}) that does not contain any Lorentz Goldstones. Then, we can interpret~(\ref{Lorentz eom}) as equations that specify the Lorentz Goldstones $\mathbbm L_\mu$. 

We therefore conclude that the inclusion of the Lorentz Goldstones does not give us anything new, so they are entirely extraneous degrees of freedom and can simply be omitted from the EFT.

\subsection{Massive}\label{Massive}

Notice that in the scalar action~(\ref{Polyakov pp}), the term involving the mass is essentially a cosmological constant-term. In the spin-1/2 action~(\ref{1/2 action}), however, there is no such mass term. The reason is that in a theory with SUSY, an ordinary cosmological constant-term of the form $\int d\tau e$ does not respect the full gauge symmetry group. In this section, we will see how a symmetry-invariant cosmological constant-term can be included in the action. 


The intuition is as follows. Suppose that the point-particle existed in a five-dimensional spacetime such that the momentum along the fifth dimension is fixed by $p^5=m$, for some constant $m>0$. Since the particle is massless, the on-shell condition is $p^\mu p_\mu + (p^5)^2=0$, or equivalently, $p^\mu p_\mu + m^2=0$. Thus, if we are only concerned with the motion of the particle along the four spacetime coordinates $x^\mu$ for $\mu=0,1,2,3$, then the particle behaves as if it has mass $m$.

Going through a procedure almost identical to that of the previous subsection, we find the effective action for a massless spin-$1/2$ particle in 5D is
\begin{equation}\begin{split} \label{massive 1/2 action 1}
S = \int d\tau \bigg(\frac{1}{2e} \dot X^\mu \dot X_\mu +\frac{i}{2} \dot \psi^\mu  \psi_\mu -\frac{i}{e} \chi \psi^\mu \dot X_\mu~~~~~~~~ \\
+ \frac{1}{2e} \dot X^5 \dot X_5 +\frac{i}{2} \dot \psi^5  \psi_5 -\frac{i}{e} \chi \psi^5 \dot X_5\bigg),
\end{split}\end{equation}
where we have defined $\mathbbm X^5(\tau,\theta) = X^5(\tau) + i\theta \psi^5(\tau)$. At this point, it is helpful to work in the ``Hamiltonian picture'' in which we include the conjugate momentum $p^5$ of $X^5$. We thus have
\begin{equation}\begin{split} 
S = \int d\tau \bigg(\frac{1}{2e} \dot X^\mu \dot X_\mu +\frac{i}{2}\dot \psi^\mu  \psi_\mu -\frac{i}{e} \chi \psi^\mu \dot X_\mu~~~~~~~~ \\
+\dot X^5 p_5 -\frac{e}{2} (p^5)^2 +\frac{i}{2} \dot \psi^5  \psi_5 -i \chi \psi^5 p_5\bigg).
\end{split}\end{equation}
Fixing $p^5=m$ for constant $m>0$ we have\footnote{Notice that we are not integrating out $p^5$; instead, we are constraining the system to have $p^5=m$.}
\begin{equation}\begin{split} 
S = \int d\tau \bigg(\frac{1}{2e} \dot X^\mu \dot X_\mu +\frac{i}{2}\dot \psi^\mu  \psi_\mu -\frac{i}{e} \chi \psi^\mu \dot X_\mu~~~~~~~~ \\
 -\frac{m^2 e}{2}+\frac{i}{2}\dot \psi^5  \psi_5 -i m \chi \psi^5 \bigg),
\end{split}\end{equation}
which agrees with the massive spin-1/2 action from~\cite{vanHolten:1995qt}.  In particular, notice that the equation of motion for $e$ is (gauge fixing $\chi=0$) 
\begin{equation}\frac{1}{e^2} \dot X^\mu \dot X_\mu + m^2 = 0,\end{equation} 
which is exactly the on-shell condition for a massive particle. 


\subsection{Coupling to gauge fields}

Oftentimes relativistic particles carry electric charge and their intrinsic quantum spin gives rise to a magnetic moment. In this subsection we will see how to couple the spin-1/2 point particles to the electromagnetic $U(1)$ gauge field. Coupling to more complicated gauge fields requires a fairly straightforward generalization. 

Suppose that $\bA_\mu(x)$ is the electromagnetic gauge field in the bulk with gauge symmetry
\begin{equation}\label{EM gauge}\bA_\mu(x)\to \bA_\mu(x) + \partial_\mu \alpha(x). \end{equation} 
 If we wish to couple it to the point particle, we must perform a pull-back to the particle's super-worldvolume by
\begin{equation}\label{SUSY pullback}\mathbb A_M(\sigma) = \frac{\partial \mathbbm X^\mu(\sigma)}{\partial\sigma^M}  \bA_\mu(\mathbbm X(\sigma)),  \end{equation} 
where $\bA_\mu(\mathbbm X(\sigma)) = \bA_\mu(X(\tau)) + i\theta\psi^\nu(\tau) \partial_\nu \bA_\mu(X(\tau))$.
Let $Q$ be the generator of the gauged $U(1)$ symmetry and let $g(\sigma)$ be the group element parameterized by the Goldstone fields. For present purposes, we will keep $g(\sigma)$ quite general. The only assumption we will make is that the worldvloume coordinates are $\sigma^M=(\tau,\theta)^M$, endowed with $\cN=1$ local SUSY. Note that since $Q$ is a symmetry of the theory, it can appear in the group element with its own Goldstone
\begin{equation}g(\sigma) = \dots e^{i \bbpi (\sigma) Q} \cdots. \end{equation} 
If $Q$ is unbroken, then $g$ may enjoy a right-acting gauge symmetry
\begin{equation}g(\sigma)\to g(\sigma)\cdot e^{i \bblambda(\sigma) Q},\end{equation} 
not to be confused with the gauge symmetry~(\ref{EM gauge}). 

The correct way to include the gauge field in Maurer-Cartan form is~\cite{Wheel}
\begin{equation} g^{-1}\big(\partial_M +i \mathbbm A_M(\sigma) Q\big) g = i \mathbbm E_M^A\cur{\cdots+ \mathbbm B_A(\sigma) Q +\cdots} ,\end{equation}
where $ \mathbbm B_A(\sigma) = \mathbbm E_A^M \mathbbm A_M(\sigma)+D_A \bbpi(\sigma)$. Then the leading-order coupling with the point-particle is 
\begin{equation}\label{EM coupling} \frac{q}{i} \int d\tau d\theta \mathbbm E~ \mathbbm B_1 (\sigma) = q\int d\tau \cur{\dot X^\mu \bA_\mu +\frac{ie}{2} \psi^\mu\psi^\nu \bF_{\mu\nu} }, \end{equation} 
where $q$ is the electric charge of the particle and $\bF_{\mu\nu} \equiv \partial_\mu \bA_\nu-\partial_\nu \bA_\mu$ is the electromagnetic field strength tensor. This coupling term agrees with the results of~\cite{vanHolten:1995qt}. Notice that the Goldstone $\bbpi$ is totally absent from this coupling.  Further, notice that the first term on the r.h.s. of~(\ref{EM coupling}) is the standard electric coupling via the pull-back of the gauge field $\bA_\mu$, while the second term describes the magnetic-spin coupling.

\section{Extended worldline SUSY}

To construct an effective action for a particle of arbitrary spin $s\in \mathbb Z/2$, we will need to endow the particle's worldline with a local extended $\cN=2s$ SUSY. In this section, we will extend the superspace formalism  of \S\ref{Worldline SUSY} to allow for multiple supercharges. To our knowledge no such superspace formalism exists in the current literature; however many common elements can be found in~\cite{Ikemori}. It will turn out that for $\cN>2$, the superspace formalism will have to be supplemented by a multiplet calculus, which is a straight-forward extension of the one presented in~\cite{vanHolten:1995qt}.

\subsection{Superspace revisited}

As before, let us begin by considering global SUSY. We define the coordinates of superspace by
\begin{equation}\sigma^M =(\tau,\theta^1,\dots,\theta^\cN)^M, \end{equation}
where $\sigma^0\equiv \tau\in \mathbb R$ and $\sigma^a\equiv \theta^a $ for $a=1,\dots,\cN$ are Grassmann odd. Suppose we have a field defined on these coordinates, $\mathbbm X(\tau,\vec\theta)$. Because $\theta^a\theta^b=-\theta^b\theta^a$, the Taylor expansion in $\theta^a$ will terminate after $\cN +1$-terms. It turns out, however, that only the first two terms are dynamical. We have
\begin{equation}\label{superfield general N}\mathbbm X(\tau,\vec\theta) = X + i\vec\theta\cdot \vec\psi+\cdots, \end{equation}
where $X(\tau)$ is a real-valued field and $\vec\psi(\tau)$ is a vector of Grassmann odd fields; thus $\mathbbm X$ is Grassmann-even. The terms subsumed by $\cdots$ in the above equation end up being non-dynamical fields whose only purpose is to allow SUSY to close off-shell.  

We define integration and differentiation with respect to the Grassmann coordinates as equivalent operations, given by
\begin{equation} \int d\theta^a \mathbbm X = \partial_a\mathbbm X = i\psi^a, \end{equation}
where $\partial_a\equiv \partial/\partial\theta^a$. 

Now we promote the global SUSY transformation to a local, gauge symmetry. Instead of introducing full superspace reparameterization invariance as we did in \S\ref{Worldline SUSY}, we will cut to the chase. We postulate that the  supervielbein takes the block-matrix form
\begin{equation}\label{high s supervielbein}
\mathbbm E ^A_M = { \begin{pmatrix}
\mathbbm  E + i\vec \theta \cdot \vec\Gamma & \Gamma^b \\
-i\theta^a &  \delta^{ab}
\end{pmatrix}_M}^A ,
\end{equation}
where $\mathbbm E = e+i \vec\theta\cdot\vec \chi$  and $\Gamma^a = \chi^a + A^{ab}\theta^b$. Here, $e$, $\chi^a$ and $A^{ab}$ are fields defined on the coordinate $\tau$.
Then the inverse supervielbein is
\begin{equation}\label{high s inversesupervielbein}
\mathbbm E ^M_A =\frac{1}{\mathbbm E} { \begin{pmatrix}
  1 & -\Gamma^b \\
i\theta^a &  \mathbbm E \delta^{ab} - i\theta^a \Gamma^b
\end{pmatrix}_A}^M .
\end{equation}
In order for the supervielbein to retain the desired form, we take our reparameterization symmetries to be a restricted subset of coordinate transformations given by $\sigma^M\to \sigma^M-\alpha^M(\sigma)$, such that
\begin{equation}\begin{split} \label{N super reparam 1}
\alpha^0(\sigma)& = \xi(\tau)+\frac{i}{\mathbbm E} \vec \theta\cdot \vec  \varepsilon(\tau), \\
\alpha^a(\sigma) & = \varepsilon^a(\tau)-\frac{i}{\mathbbm E}\vec\theta\cdot\vec\varepsilon(\tau) \Gamma^a+\theta^b \beta^{ab}(\tau),
\end{split}\end{equation}
where $\xi(\tau)$ and $\beta^{ab}(\tau)=-\beta^{ba}(\tau)$ are infinitesimal and real-valued, while $\varepsilon^a(\tau)$ are infinitesimal and Grassmann-odd. Under this coordinate transformation, the component fields of the supervielbein transform as
\begin{equation}\begin{split}\label{N super reparam 2}
\delta e& = \partial_\tau(\xi e ) + 2 i \vec\varepsilon\cdot \vec \chi, \\
\delta\chi^a &= \partial_\tau(\xi \chi^a) +\dot \varepsilon^a + A^{ab} \epsilon^b -\beta^{ab}\chi^b, \\
\delta A^{ab} &= \partial_\tau(\xi A^{ab}) + \dot \beta^{ab} +\beta^{ac} A^{cb} - \beta^{bc}A^{ca}. 
\end{split} \end{equation} 
We can therefore interpret $e$ as the worldline einbein, $\chi^a$ as the gravitinos and $A^{ab}$ as an $O(\cN)$ gauge field. Notice that in the case $\cN=1$, $A^{ab}$ vanishes, which is why we did not encounter it in the spin-1/2 case. 

The covariant derivatives are given by $D_A\equiv \mathbbm E_A^M\partial_M$. Explicitly, we have
\begin{equation} D_0 = \frac{1}{\mathbbm E} (\partial_\tau -\vec\Gamma \cdot \vec\partial),~~~~~D_a \equiv i\theta^a D_0 +\partial_a . \end{equation} 
When acting on the scalar field $\mathbbm X$, we have
\begin{equation}\begin{split}\label{covariant derivatives cN} D_0 \mathbbm X &= \frac{1}{\mathbbm E} (z+i \vec\theta\cdot\vec\zeta) \cdots, \\
D_a \mathbbm X& = \frac{i}{\mathbbm E} \theta^a \cur{z +i\vec\theta\cdot \vec\zeta}+ i\psi^a +\cdots,
\end{split}\end{equation} 
where $ z\equiv  \dot X-i\vec \chi\cdot \vec\psi$ and $\zeta^a \equiv \dot \psi^a +A^{ab}\psi^b$ and $\cdots$ denote terms with non-dynamical fields. Finally, the invariant integration measure is $\int d\tau d^\cN \theta \mathbbm ~\mathbbm E$.



\subsection{Method of multiplets}

The construction of a SUSY-invariant action using the superspace formalism by integrating a Lagrangian over the whole of superspace for $\cN>2$ is still an unsolved problem. In fact, there is good reason to believe it impossible, which we will address in a later section. Further, even if we were to accomplish such a feat, the resulting action would contain a potentially very large number of auxiliary fields that are entirely non-dynamical and would only serve to ensure that the SUSY transformations close off-shell. We could then integrate out such non-dynamical fields to obtain a simpler action that would enjoy SUSY only on-shell. It is the aim of this subsection to explain how to directly construct the on-shell SUSY action without worrying about the non-dynamical fields. We term this approach the method of multiplets. 

Our multiplet approach will enable us to construct off-shell SUSY-invariant actions for global SUSY. Only at the end will we gauge this symmetry to obtain the desired local on-shell SUSY-invariant action. We begin by defining bosonic and fermionic multiplets. A bosonic multiplet is an ordered pair
\begin{equation}\Sigma = (X,\vec\psi),\end{equation}
where $X$ is Grassmann even and $\vec\psi$ is a Grassmann-odd $\cN$-component vector. Under an infinitesimal global SUSY transformation, we have
\begin{equation} \delta X= \xi \dot X+i\vec \varepsilon\cdot \vec\psi,~~~~~\delta\psi^a =\xi\dot\psi^a +\varepsilon^a \dot X -\beta^{ab}\psi^b,\end{equation}
where $\xi$ is an infinitesimal real constant and $\vec\varepsilon$ and $\beta^{ab}=-\beta^{ba}$ are Grassmann-odd constants. Next, we define a fermionic multiplet as an ordered pair 
\begin{equation}\Phi = (\vec f, b), \end{equation}
where $\vec f$ s a Grassmann-odd $\cN$-component vector and $b$ Grassmann even. The components transform by
\begin{equation} \delta f^a = \xi \dot f^a +i\varepsilon^a b-\beta^{ab} f^b,~~~~~\delta b =\xi \dot b +\vec\varepsilon\cdot\dot {\vec f} .  \end{equation} 

We can add and multiply these multiplets together. The rule for addition is simply component-wise addition,
\begin{equation}\begin{split}\label{mult add} (X_1,\vec \psi_1)+ (X_2,\vec \psi_2) &= (X_1+X_2, \vec \psi_1 +\vec \psi_2), \\
(\vec f_1,b_1)+ (\vec f_2,b_2) &= (\vec f_1+\vec f_2,b_1+b_2). 
\end{split}\end{equation}
We are not permitted to add a bosonic multiplet to a fermionic multiplet. Next, we have the multiplication rules
\begin{equation}\begin{split}\label{mult mult} (X_1,\vec \psi_1)\times (X_2,\vec \psi_2) &= (X_1 X_2, X_1 \vec \psi_2 +X_2 \vec \psi_1), \\
(\vec f_1,b_1)\times (\vec f_2,b_2) &= (\vec f_1\cdot\vec f_2, b_1\vec f_2-b_2\vec f_1 ), \\
(X,\vec \psi)\times(\vec f, b) &= (X\vec f,Xb + \vec \psi\cdot \vec f ).  \end{split} \end{equation}
Notice that the first two products yield bosonic multiplets and the last product yields a fermionic multiplet. 

Additionally, we can take derivatives of these multiplets. Acting on a bosonic multiplet  we have
\begin{equation}\label{bos mult deriv} D_0(X,\vec\psi) = \cur{\dot X,\dot{\vec\psi}} ,~~~~~ D_a (X,\vec\psi) = \cur{i\psi^a,\dot X}. \end{equation}
It can be checked that $D_0 (X,\vec\psi)$ and $D_a(X,\vec\psi)$ are, respectively, bosonic and fermionic multiplets. 
Acting on a fermionic multiplet, we have
\begin{equation} D_0(\vec f, b) = \cur{\dot {\vec f},\dot{b}} ,~~~~~ D_a (\vec f,b) = \cur{ib, \dot{\vec f}}. \end{equation} 
It can be checked that $D_0 (\vec f, b)$ and $D_a(\vec f, b)$ are, respectively, fermionic and bosonic multiplets. Oftentimes, we will use the compact notation $D_A$ for $A=0,\dots,\cN$. 
Lastly, given the fermionic multiplet $\Phi = (\vec f, b)$, the global SUSY-invariant integral is 
\begin{equation}\label{mult integrate}\int_\text{SUSY} \Phi \equiv \int d\tau ~b. \end{equation}
We therefore see that the aim of the coset construction will be to construct a symmetry-invariant fermionic multiplet that will then be integrated according to the above rule. After the action is constructed in this manner, we can include the worldline gauge fields to promote the global SUSY to a local, gauge symmetry. 


\section{Higher-spin point-particles}

The symmetry-breaking pattern of the Poincar\'e group for massless particles is the same for all values of spin~$s$, namely the unbroken generators are~(\ref{unbroken massless}) and the broken generators are~(\ref{broken massless}). Thus, the only possible distinguishing features among the effective actions for particles of differing spins are the choices of worldvolume and gauge symmetries. It has been demonstrated in~\cite{Ikemori} that the EFT for a particle of spin $s$ enjoys  $\pazocal N=2s$ SUSY. At this stage, we will split the problem into two pieces. The first deals with spin 1 particles, corresponding to $\cN=2$. The effective action will be defined in terms of an integral over superspace and will thus lead to an off-shell SUSY-invariant action. The second will be valid for all $\cN\geq 1$, but will instead rely on the multiplet formalism and as a result will yield local on-shell SUSY-invariant actions. 

\subsection{Spin-1 point particles}

We define our coset on the superspace coordinates $\sigma^M=(\tau,\vec\theta)^M$ for $\vec\theta = (\theta^1,\theta^2)$. We take the unbroken translation generator $P_u$ as an opportunity  to impose the super-reparameterization invariance~(\ref{N super reparam 1}) and~(\ref{N super reparam 2}), where the supervielbein is given by~(\ref{high s supervielbein}). Then, as in the spin-1/2 case, we take the unbroken little group generators $\cJ_i$ as an opportunity to impose a gauge invariance that can be used to fix $\bbtheta^i=0$. 

With this gauge-fixing condition the most general group element is
\begin{equation}g(\sigma) =  e^{i \mathbbm X^\alpha(\sigma) P_\alpha+ i \mathbbm X^m (\sigma) P_m}  e^{i \bbeta^i(\sigma) \cK_i}, \end{equation}
where $\alpha=u,v$ and $m=1,2$. 
The Maurer-Cartan form is 
\begin{equation}\begin{split} g^{-1} \partial_M g = i \mathbbm E^A_M \big (\nabla_A \mathbbm X^\alpha P_\alpha&+\nabla_A\mathbbm X^m P_m \\&+ \nabla_A\bbeta^i \cK_i +\BOmega^i_A \cJ_i \big),\end{split} \end{equation} 
where the covariant derivatives of $\mathbbm X^\mu$ are 
\begin{equation}\begin{split}
\nabla_A\mathbbm X^\alpha & = \mathbbm E_A^M \partial_M \mathbbm X^\mu {\BLambda _\mu}^\alpha ,\\
\nabla_A\mathbbm X^m & = \mathbbm E_A^M \partial_M \mathbbm X^\mu {\BLambda _\mu}^m ,
\end{split} \end{equation}
such that ${\BLambda^\mu}_\nu = (e^{i \bbeta^i(\sigma) \cK_i} {)^\mu}_\nu$.   The precise forms of $\nabla_A \bbeta^i $ and $\BOmega^i _A$ will not be important for the construction of the leading-order action. 

We impose on-shell IH constraints 
\begin{equation}\label{on shell IH}0= \nabla_0\mathbbm X^v ,~~~~~0=\nabla_0 \mathbbm X^m, \end{equation} 
which allow us to remove $\bbeta^i$ from the effective action. The leading-order effective action for $\cN=2$ is therefore 
\begin{equation}S = \frac{1}{4} \int d\tau d^2\theta \mathbbm E~ \epsilon^{ab} \nabla_a\mathbbm X^\mu \nabla_b \mathbbm X_\mu ,\end{equation}
which can be immediately simplified to 
\begin{equation}S = \frac{1}{4} \int d\tau d^2\theta \mathbbm E~ \epsilon^{ab} D_a\mathbbm X^\mu D_b \mathbbm X_\mu ,\end{equation}
where $\epsilon^{ab}$ is the totally antisymmetric tensor such that $\epsilon^{12}=-\epsilon^{21}=1$ and $\epsilon^{11}=\epsilon^{22}=0$. We can now expand the superfield $\mathbbm X^\mu$ in powers of $\vec\theta$, yielding
\begin{equation} \mathbbm X^\mu = X^\mu + i\theta^a \psi^{a\mu} + \frac{i}{2} \epsilon^{ab}\theta^a\theta^b F^\mu.  \end{equation}
The integral over $ d^2\theta$ can now be explicitly evaluated, yielding an action defined on the worldline coordinate $\tau$ by
\begin{equation}\label{spin-1 action}\begin{split}
S = \int d\tau \bigg( \frac{1}{2e} (\dot X - i \vec \chi\cdot\vec \psi) ^2  +\frac{i}{2} \dot{\vec \psi}^\mu \cdot  {\vec \psi}_\mu  +\frac{e}{2} F^2 & \\- \frac{i}{2} A^{ab}\psi^{\mu a}\psi_\mu^b &\bigg). 
\end{split} \end{equation}
Notice that the field $F^\mu$ appears with no derivatives and that its equation of motion is simply $F^\mu=0$. We can thus integrate it out at no cost, except the resulting action will only be invariant under SUSY on-shell.

\subsection{Arbitrary-spin point particles}

As mentioned earlier, for particles of spin $s>1$---which correspond to $\cN>2$---we must admit defeat, at least in part, if we wish to construct a SUSY-invariant action using the superspace formalism. The reason is that no superspace integral can give rise to the desired invariant action. To see why, we propose the following power-counting argument. We let $\mu$ count powers of energy; e.g. $[\partial_\tau] =  [d\tau]^{-1} = \mu^1$. Begin by noticing that the actions we have calculated so far,~(\ref{Polyakov pp},\ref{1/2 action},\ref{spin-1 action}) all contain a term of the form
\begin{equation}S\supset \int d\tau \frac{1}{2 e } \dot X^2+\cdots. \end{equation}
By assuming $S$ scales as $\mu^0$, we therefore have the scaling of $X$, and hence $\mathbbm X$, namely $[\mathbbm X] = \mu^{-1}$. Further, notice that the commutator of two super-translations give rise to one temporal translation, so $[D_\theta] = [D_0]^{1/2} = \mu^{1/2}$. Moreover, integration and differentiation with respect to Grassmann numbers are equivalent operations, so $[d\theta]= \mu^{1/2}$. Lastly, our action should involve two factors of $\mathbbm X$ and two covariant derivatives $D$ and $D'$, which can be either $D_0$ or $D_\theta$. Then the action scales like 
\begin{equation}[S] =[  d\tau][ d^\cN \theta] [D][ \mathbbm X][ D'][ \mathbbm X] = [D][D'] \mu^{s-3},\end{equation}
where we have used the fact that $\cN=2s$. 
But also, the action ought to scale like $\mu^0$, so we have $[D][D'] = \mu^{3-s}$. But since $D$ and $D'$ must scale as either $\mu^1$ or $\mu^{1/2}$, this equation can only be satisfied if $s\leq 1$. In particular, for $s=0$, we have $D=D'=D_0$; for $s=1/2$, we have $D=D_0$ and $D'=D_\theta$; and for $s=1$, we have $D=D'=D_\theta$. \\

We now use the multiplet formalism to construct actions invariant under global SUSY for arbitrary $\cN$ and hence arbitrary spin. Let $\Sigma^\mu = (X^\mu,\vec\psi^\mu)$ and $\upeta^i = (\eta^i,\vec\phi^i)$ be bosonic multiplets. We take the unbroken generators $\cJ_i$ as an invitation to impose gauge symmetries that allow us to remove the Goldstones associated with $\cJ_i$. Then, we can express the most general group element as the SUSY multiplet\footnote{Notice that the addition and multiplication rules~(\ref{mult add}) and~(\ref{mult mult}) enable us to define arbitrary functions of the multiplets by employing a Taylor expansion. } 
\begin{equation}g = e^{i\Sigma^\alpha P_\alpha}e^{i\Sigma^m P_m} e^{i\upeta^i \cK_i}.\end{equation}
Using the SUSY covariant derivative (not to be confused with the covariant derivative generated by the coset) defined by~(\ref{bos mult deriv}), we can compute the Maurer-Cartan form by
\begin{equation} g^{-1} D_A g = i\cur{\nabla_A \Sigma^\alpha P_\alpha+\nabla_A \Sigma^m P_m +\nabla_A \upeta^i \cK_i+\Upomega^i \cJ_i}, \end{equation}
where the covariant derivatives of $\Sigma^\mu$ generated by the coset are
\begin{equation}\begin{split} \nabla_A \Sigma^\alpha &=D_A \Sigma^\mu {\Uplambda_\mu}^\alpha,\\
\nabla_A \Sigma^m &=D_A \Sigma^\mu {\Uplambda_\mu}^m, \end{split} \end{equation}
where ${\Uplambda^\mu}_\nu = (e^{i\upeta^i\cK_i}{)^\mu}_\nu$. The precise form of the other terms in the Maurer-Cartan form will not be important for the construction of the leading-order action. To remove the extraneous Goldstones $\upeta^i$, we impose the on-shell IH constraints\footnote{It should be noted that these are only legitimate on-shell IH constraints because we are going to ultimately gauge the worldline SUSY. If we were to keep the SUSY global, then there would be no constraints imposed by gauge fields and the equations of motion arising from boost Goldstones would in fact be extremely important.} 
\begin{equation}0=\nabla_0 \Sigma^v,~~~~~ 0 = \nabla_0 \Sigma^m.\end{equation}

To construct the effective action, we must build a symmetry-invariant fermionic multiplet, which can then be integrated according to the prescription~(\ref{mult integrate}). We see that at leading order, the only such option is  $\Phi \equiv \nabla_0\Sigma^\mu \nabla_a \Sigma_\mu$, which can be immediately simplified to $\Phi = D_0\Sigma^\mu D_a \Sigma_\mu$. 
We then have the leading-order action 
\begin{equation}S = \int_\text{SUSY}\Phi = \int d\tau \cur{\frac{1}{2} \dot X^2 + \frac{i}{2} \dot{\vec\psi}^\mu\cdot\vec\psi_\mu}. \end{equation}
We must now gauge the $\cN=2s$ extended worldline SUSY. In particular, we want an action that in invariant, up to terms proportional to the equations of motion, under the infinitesimal local transformations~(\ref{N super reparam 2}) and the on-shell SUSY transformations 
\begin{equation}\begin{split} \delta X^\mu &= \xi \dot X^\mu + i \vec\varepsilon \cdot \vec\psi^\mu,\\
\delta\psi^a &=\xi \dot \psi^{a\mu} + \varepsilon^a \frac{1}{e} \cur{\dot X^\mu - i\vec\chi \cdot \vec\psi^\mu} -\beta^{ab} \psi^{b\mu}, \end{split} \end{equation} 
where $\xi$, $\vec\varepsilon$, and $\beta^{ab}$ are now generic infinitesimal functions of $\tau$. 
In particular, we have
\begin{equation}S =  \int d\tau \cur{\frac{1}{2e} z^2 + \frac{i}{2} \vec\zeta^\mu \cdot\vec\psi_\mu}, \end{equation}
where where $ z\equiv  \dot X-i\vec \chi\cdot \vec\psi$ and $\zeta^a \equiv \dot \psi^a +A^{ab}\psi^b$.
Equivalently, the expanded form of the action is
\begin{equation}\label{higher s action}\begin{split}
S = \int d\tau \bigg( \frac{1}{2e} (\dot X - i \vec \chi\cdot\vec \psi) ^2  +\frac{i}{2} \dot{\vec \psi}^\mu \cdot  {\vec \psi}_\mu    - \frac{i}{2} A^{ab}\psi^{\mu a}\psi_\mu^b\bigg). 
\end{split} \end{equation}
Comparing with~\cite{Howe:1988ft}, we find that we have successfully reproduced the action for on-shell extended local SUSY.

If we wish to include a mass-term in this action, we can follow the steps of \S\ref{Spin-1/2 point-particles}\ref{Massive} and arrive at the following additional terms in the action
\begin{equation}\label{massive higher s action}\begin{split}
S_\text{mass} = \int d\tau \bigg(-\frac{m^2 e}{2}+\frac{i}{2} \dot{ \vec \psi}^5\cdot   {\vec \psi}_5 - i m \vec\chi\cdot\vec\psi^5& \\-\frac{i}{2} A^{ab}\psi^{5a}\psi_5^{b}& \bigg). 
\end{split} \end{equation}
Thus, the full action for the massive higher-spin point particle is the sum of~(\ref{higher s action}) and~(\ref{massive higher s action}).

We close this section by coupling higher-spin point-particles to gauge fields using the multiplet formalism. Let $\bA_\mu(x)$ be the electromagnetic $U(1)$ gauge field. In order for this field to appear in the Maurer-Cartan form, we must first convert it into a SUSY multiplet. Inspired by~(\ref{SUSY pullback}), we define the multiplet pullback by
\begin{equation}\cA_A = D_A\Sigma^\mu \bA_\mu(\Sigma) , \end{equation} 
where $\bA_\mu(\Sigma) = (\bA_\mu(X(\tau)), \vec \psi^\nu(\tau) \partial_\nu \bA_\mu(X(\tau)).$
We can now insert this multiplet pullback gauge field into the Maurer-Cartan form by
\begin{equation}g^{-1}\cur{D_A+i\cA_A Q} g ,\end{equation} 
where $Q$ generates the $U(1)$ symmetry that is gauged by $\bA_\mu$. Letting $\uppi$ be the Goldstone multiplet field corresponding to $Q$, we have 
\begin{equation}g^{-1}\cur{D_A+i\cA_A Q} g = \cdots+ \cB_A Q+\cdots,\end{equation} 
where $\cB_A = \cA_A +i D_A\uppi$. 

To construct the leading-order coupling term, we must integrate a fermionic multiplet. The only option at leading-order is $\cB_a$, so we have the global SUSY-invariant term
\begin{equation}q \int_\text{SUSY} \cB_a = q\int d\tau \cur{\dot X^\mu \bA_\mu +\frac{i}{2} \vec \psi^\mu\cdot\vec\psi^\nu \bF_{\mu\nu} }, \end{equation} 
where $q$ is the electric charge of the point-particle and $\bF_{\mu\nu} = \partial_\mu \bA_\nu-\partial_\nu\bA_\mu$ is the electromagnetic field strength.
Finally, gauging the worldine SUSY gives us 
\begin{equation} q\int d\tau \cur{\dot X^\mu \bA_\mu +\frac{i e}{2} \vec \psi^\mu\cdot\vec\psi^\nu \bF_{\mu\nu} }. \end{equation} 
 Notice that for $\cN=1$, we recover the spin-$1/2$ coupling term~(\ref{EM coupling}) that we derived using the superspace formalism. 

\section{Quantization} 
After employing our novel coset construction to formulate actions with worldline SUSY, we now wish to demonstrate that they in fact describe point particles with intrinsic quantum spin. By quantizing the action with $\cN=2s\in \mathbb Z$ worldline SUSY, we will find that the resulting wave function describes a point-particle of spin~$s$. We will follow closely the procedure shown in \cite{Howe:1988ft}. 

Starting with equation (\ref{higher s action}), we find it convenient to express the action in the ``Hamiltonian form,''
\begin{equation}\begin{split}
S = \int d\tau \bigg( p^{\mu}\dot{X}_{\mu} &- \frac{1}{2}ep^{\mu}p_{\mu} - i \vec{\chi}\cdot\vec{\psi}^{\mu}p_{\mu}  \\&+\frac{i}{2} \dot{\vec {\psi}}^\mu \cdot  {\vec \psi}_\mu - \frac{i}{2} A^{ab}\psi^a\psi^{b} \bigg).
\end{split}\end{equation}
It is then clear from the equations of motion for $e$, $\chi^a$ and $A^{ab}$ that 
\begin{align}{\label{constraints for quantization}}
p^{\mu}p_{\mu} &= 0 \nonumber, \\
p^{\mu}\psi^a_{\mu}&=0 \nonumber, \\
(\psi^{\mu})_{[a}(\psi_{\mu})_{b]} &=0.
\end{align}
We then impose the (anti-)commutation relations 
\begin{equation}
[ X^{\mu}, \; p_{\nu} ] = i\delta^{\mu}_{\nu},~~~~~ \{ \psi^{\mu}_a,\; \psi^{\nu}_b \}=\eta^{\mu\nu}\delta_{ab}.
\end{equation}
To facilitate the quantization procedure, it is convenient to work with the usual identification  $p_{\mu} = -i\partial/\partial X^{\mu}$. To realize the second anti-commutation relation, we work in the representation 
\begin{equation}
\psi^{\mu}_a = \frac{1}{\sqrt{2}} \underbrace{\gamma_5\otimes\cdots\otimes\gamma_5}_{a-1}\otimes\gamma^{\mu}\otimes\underbrace{\mathbb{1}\otimes\cdots\otimes\mathbb{1}}_{\cN-1},
\end{equation}
where we have used the Dirac $\gamma$ matrices, which satisfy $\{\gamma^\mu,\gamma^\nu\} = 2\eta^{\mu\nu}$. Now, assuming $a<b$, it is straightforward to determine that
\begin{equation}
\begin{split}
(\psi^{\mu})_{[a}(\psi_{\mu})_{b]} = -\frac{1}{2}\underbrace{\mathbb{1}\otimes\cdots\otimes\mathbb{1}}_{a-1}\otimes\gamma_5\gamma^{\mu}\otimes\underbrace{\gamma_5\otimes\cdots\otimes\gamma_5}_{b-a-1} \\
\otimes\gamma_{\mu}\otimes\underbrace{\mathbb{1}\otimes\cdots\otimes\mathbb{1}}_{\cN-b}.
\end{split}
\end{equation}
The constraint equations coming from equations (\ref{constraints for quantization}) for a spinor $\Psi_{\alpha_{1}\cdots\alpha_{\cN}}$ are, respectively,
\begin{align}
\square \Psi_{\alpha_{1}\cdots\alpha_{\cN}} =&0 \nonumber, \\
\slashed{\partial}_{\alpha_{a}}^{\;\;\; \beta_{a}} \Psi_{\alpha_{1}\cdots\beta_{a}\cdots\alpha_{\cN}} =&0 \nonumber, \\
(\gamma^{\mu})_{\alpha_{a}}^{\;\;\; \beta_{a}}(\gamma_{\mu})_{\alpha_{b}}^{\;\;\; \beta_{b}}\Psi_{\alpha_{1}\cdots\beta_{a}\cdots\beta_{b}\cdots\alpha_{\cN}} =&0.
\end{align}
Following the procedures in \cite{Howe:1988ft}, these constraints will finally give, in the $SL(2,\mathbb{C})$ notation, 
\begin{equation}
\partial_{\beta_{a}}^{\dot{\alpha}_a}\Psi_{\dot{\alpha}_{1}\cdots \dot{\alpha}_{a}\cdots \dot{\alpha}_{N}} =0,~~~~~ \square \Psi_{\alpha_{1}\cdots \alpha_{\cN}} =0. 
\end{equation}
But these are just the conditions that $\Psi_{\alpha_{1}\cdots \alpha_{N}}$ be the wave function for a relativistic particle of spin $s=\cN/2$. We therefore conclude that our action (\ref{higher s action}) indeed describes a spin-$\cN/2$ point particle.

\section{Massless spin-0 point-particles}

When we constructed the EFTs for the massive boson, the resulting IH constraints could be solved directly for the Lorentz Goldstones, $\eta^i$ in terms of the translation Goldstones, $X^\mu$. However, when dealing with massless particles, the IH constraints over-constrained the translation Goldstones and we were forced to interpret them as on-shell IH constraints. In this section we will explore in more detail the relationship between Lorentz Goldstones and massless particles.  

\subsection{Without external einbein}
Our aim is to construct an EFT for massless scalars without the introduction of an external einbein. The SSB pattern is identical to that of the spin-1/2 (and higher) massless particles. Explicitly, the unbroken generators are~(\ref{unbroken massless}) and the broken generators are~(\ref{broken massless}). The most generic group element is
\begin{equation}g(\tau) =  e^{i  X^\alpha(\tau) P_\alpha+ i  X^m (\tau) P_m}  e^{i \eta^i(\tau) \cK_i} e^{i\vartheta^i(\tau) \cJ_i}, \end{equation}
where $\alpha=u,v$ and $m=1,2$. 
We impose full reparameterization symmetry $\delta \tau = -\xi(\tau)$ as well as a full $\cJ_i$ gauge symmetry
\begin{equation}g\to g\cdot e^{i\lambda^i(\tau) \cJ_i},\end{equation}
which allows us to gauge-fix $\vartheta^i=0$. With this gauge-fixing condition, the Maurer-Cartan form is
\begin{equation}g^{-1} \partial_\tau g = i E(P_u + \nabla X^v P_v +\nabla X^m P_m +\nabla\eta^i ) +\Omega^i,\end{equation} 
where the einbein is given by $E = \dot X^\mu {\Lambda_\mu}^u$ and ${\Lambda^\mu}_\nu = (e^{i\eta^i(\tau)\cK_i}{)^\mu}_\nu$. The covariant derivatives and spin connection will not appear at leading order in our EFT.

One might be tempted to impose the IH constraints $\nabla X^v =0$ and $\nabla X^m=0$, but this would imply that $\dot X^\mu$ is null. Further, since there are no external gauge fields that might constrain $\dot X^\mu$ to be null, these constraints do not admit an interpretation as on-shell IH constraints. Thus, we are forced to include the Lorentz Goldstones $\eta^i$ in our EFT.

The leading-order action is then
\begin{equation}\label{massless boson no einbein} S = \int d\tau E = \int d\tau \dot X^\mu L_\mu,\end{equation} 
where $L_\mu\equiv {\Lambda_\mu}^u$. Importantly, $L_\mu$ is constrained to be a null vector, that is $L^2=0$. 

The equations of motion found by varying $L_\mu$ and $X^\mu$ are respectively 
\begin{equation} \dot X^\mu \propto L^\mu,~~~~~ \dot L_\mu = 0.  \end{equation}
Since $L^2=0$ the first equation tells us $\dot X^2=0$, meaning that the particle travels along a null-trajectory. Then, the second equation tells us that $\ddot X^\mu=0$, so the particle does not accelerate. This is exactly what we should expect of a massless particle. Thus, we have an effective action for a massless boson that does not include any external eibein at all; the price we pay is the inclusion of Lorentz Goldstones.

\subsection{With external einbein}

We now consider what happens if we include an external einbein $e$ such that worldline diffeomorphism symmetry acts by
\begin{equation}\delta\tau = -\xi(\tau), ~~~~~\delta e =\partial_\tau (e\xi). \end{equation}
As in the previous subsection we gauge-fix $\vartheta^i=0$. The Maurer-Cartan form is
\begin{equation}g^{-1} \partial_\tau g = i e( \nabla X^\alpha P_\alpha +\nabla X^m P_m +\nabla\eta^i ) +\Omega^i,\end{equation} 
where $\alpha=u,v$ and $m=1,2$. Since we have an external eibein, we should expect that on-shell IH constraints are permitted, and this is indeed the case. To be explicit about how these on-shell IH constraints can be used, we will wait to impose any IH constraints and keep $\eta^i$ in the action. 

The leading-order action is 
\begin{equation}\label{massless boson with einbein} S= - \int d\tau e \cur{ \nabla X^\mu \nabla X_\mu +C \nabla X^u+m^2},  \end{equation}  
for some constants $C$ and $m^2$. Define the field $L_\mu\equiv C{\Lambda_\mu}^u$, which is subject to the constraint $L^2=0$. Then our action becomes
\begin{equation} S=- \int d\tau  \cur{\frac{1}{e} \dot X^\mu \dot X_\mu +  \dot X^\mu L_\mu - m^2 e},  \end{equation}  
The constraint equation coming from $e$ is
\begin{equation} \frac{1}{e^2} \dot X^2= -m^2 \end{equation}
and the constraint equations coming from $L_\mu$ are
\begin{equation}\label{L eom}\dot X^\mu\propto L_\mu.\end{equation} 
Upon squaring~(\ref{L eom}), we find that $\dot X^2=0$, which then forces $m^2=0$ on self-consistency grounds. 

Thus, the effect of the on-shell IH constraint is to fix $m^2=0$. After we impose this constraint on $m^2$, we can simply drop the Lorentz Goldstones from our action. Thus the leading-order action is
\begin{equation} S=- \int d\tau  \frac{1}{e} \dot X^\mu \dot X_\mu, \end{equation}  
which is the standard action for a massless spin-0 point-particle. 

Finally, it is worth noting that ordinary IH constraints allow one to solve for the extraneous Goldstone in terms of other Goldstones; however, dynamical IH constraints~\cite{Ira, Ira 2} instead serve to impose operator constraints on the terms of the action. Notice that the equations of motion for $L_\mu$ in the action~(\ref{massless boson no einbein}) impose the operator constraint $\dot X^2=0$. We therefore may conceive of the equations of motion for $L_\mu$ as imposing dynamical IH constraints. Similarly when we impose on-shell IH constraints in the action~(\ref{massless boson with einbein}) (which is equivalent to computing the equations of motion for $L_\mu$) we similarly find a constraint equation that forces $m^2=0$. Thus on-shell IH constraints bear a striking resemblance to dynamical IH constraints. 

\section{Spinning particles}

So far, we have been dealing with point-particles, that is particles with no spatial extent. Now we turn our attention to particles that have spatial extent. As a result, there may exist angular momentum in the form of quantum spin or classical rotation. 

\subsection{Spinning spin-less particles}

Consider a spin-0 particle with spatial extent. Since we are keen on implementing our new coset construction, we will focus on constructing the Polyakov-type action with an intrinsic worldline einbein. The Nambu-Goto-type action for a spinning spin-0 particle can be constructed with the ordinary coset construction; interested readers may consult~\cite{Wheel} for such a derivation. 

As before, the symmetry group is Poincar\'e alone, so the most general group element is given by~(\ref{Polyakov group elt}). We impose worldline-reparameterization symmetry on the coordinate $\tau$ given by~(\ref{world line diff}), where once again $e(\tau)$ is the einbein. Because the particle has finite spatial extent, rotations are spontaneously broken, so we cannot impose a local rotational gauge symmetry of the form~(\ref{local rotation gauge}). Thus, the rotation Goldstones cannot be gauge-fixed to zero. 

We have supposed that our particle has some sort of finite extent, but we have so far been vague about its shape. At the level of the coset construction, what distinguishes a lumpy object from a sphere or a cylinder? The answer is symmetry: a sphere has an internal $SO(3)$ symmetry, a cylinder has an internal $SO(2)\times \mathbb Z_2$ symmetry, and a lumpy object has no internal symmetry. To treat all of these possibilities (and more) simultaneously, we suppose that the symmetry group of the object is $\cS\subset SO(3)$. Then we impose the rigid symmetry
\begin{equation}\label{rotation symm} g\to g\cdot S, \end{equation} 
where $S\in \cS$ is constant.\footnote{This right-acting rigid rotation symmetry is an example of~(\ref{rigid gauge trans}).} 

We may now compute the Maurer-Cartan form
\begin{equation}
g^{-1} \partial_\tau g = i e(\nabla X^\mu P_\mu + \nabla\eta^i K_i +\nabla\vartheta^i J_i),
\end{equation}
where the covariant derivatives are
\begin{equation}\begin{split}
\nabla X^0 & = e^{-1} \dot X^\nu {\Lambda_\nu}^0 ,\\
\nabla X^i & = e^{-1} \dot X^\nu {\Lambda_\nu}^j R^{ij} ,\\
\nabla \eta^i & = e^{-1} (\Lambda^{-1} \partial_\tau \Lambda)^{0j} R^{ji},\\
\nabla\vartheta^i & = \frac{e^{-1}}{2} \epsilon^{ijk} ((\Lambda R)^{-1} \partial_\tau (\Lambda R) )^{jk}, 
\end{split} \end{equation}
where ${\Lambda^\mu}_\nu\equiv (e^{i\eta^i(\tau)K_i}{)^\mu}_\nu$ and $R^{ij}\equiv (e^{i\vartheta^i(\tau) J_i})^{ij}$. 
We impose the IH constrains $\nabla X^i=0$, which can be solved to give~(\ref{spin0 IH}), but now since we have to contend with the rotation Goldstones, it is profitable to dig deeper into these IH constraints. In particular, defining the orthonormal basis of vectors 
\begin{equation} n^{(i)}_\mu = {\Lambda_\mu}^j R^{ji},~~~~~ u^\mu = {\Lambda^\mu}_0, \end{equation} 
the IH constraints merely state that $e^{-1} \dot X^\mu$ is orthogonal to $n^{i}_\mu$, meaning that $e^{-1} \dot X^\mu$ is parallel to $u^\mu$. We can thus identify $u^\mu$ as the four-velocity of the particle. Therefore, the matrix $R^{ij}$ encodes information about the rotation of the particle in its own inertial rest frame. 

With the boost Goldstones successfully eliminated, $\nabla \vartheta^i$ is now expressible entirely in terms of $X^\mu$ and $\vartheta^i$ alone. Recalling~(\ref{recalling}), the leading-order action is given by 
\begin{equation}\begin{split}\label{Polyakov spinning}
S =- \int d\tau \cur{ \frac{1}{e} \dot X^\mu\dot X_\mu - e m^2 -\frac{e}{2} I_{ij} \nabla \vartheta^i\nabla\vartheta^j},\end{split}\end{equation}
where $I_{ij}$ is the moment of inertia tensor. Notice that the internal symmetry group $\cS$ determines the symmetries of~$I_{ij}$.

\subsection{Spinning spinning particles}

We now consider the case of a particle with spin $s= 1/2$ and finite spatial extent. Such a particle, in addition to spin-angular momentum, may now have `orbital' angular momentum, or classically, it may rotate in space. We are primarily interested in massive particles as the concept of rotation is most evident when a particle has an inertial rest frame. We thus take the global symmetry group to be Poincar\'e with the addition of a $U(1)$ symmetry generated by $P_5$. We can think of $P_5$ as generating translations along a compactified dimension. As we have done previously for massive particles with spin, we will ultimately constrain the momentum along the 5-direction by $p^5=m$. Then we can interpret $m$ as the mass. 

The additional symmetries we impose are the usual $\cN=2s$ local SUSY and the internal rotation symmetry~(\ref{rotation symm}). The most general group element is
\begin{equation}g(\sigma) =  e^{i \mathbbm X^\mu(\sigma) \bar P_\mu} e^{i \mathbbm X^5(\sigma) P_5} e^{i \bbeta^i(\sigma) K_i} e^{i\bbtheta^i(\sigma) J_i}, \end{equation}
where $\bar P_\mu \equiv P_\mu - \delta_\mu^0 P_5$. 
The resulting Maurer-Cartan form is
\begin{equation}\begin{split} g^{-1}\partial_M g = i (\mathbbm E^{-1})^A_M (\nabla_A \mathbbm X^\mu P_\mu +\nabla_A \mathbbm X^5 ~~~~~~~~~~\\ + \nabla_A\bbeta^i K_i +\nabla_A \bbtheta^i J_i ),\end{split}\end{equation} 
where the covariant derivatives  are 
\begin{equation}\begin{split}
\nabla_A\mathbbm X^0 & = \mathbbm E_A^M  \partial_M \mathbbm X^\nu {\BLambda_\nu}^0 ,\\
\nabla_A\mathbbm X^i & = \mathbbm E_A^M  \partial_M \mathbbm X^\nu {\BLambda_\nu}^j \mathbbm R^{ji} ,\\
\nabla_A\mathbbm X^5 & = \mathbbm E_A^M  \partial_M \mathbbm X^5 +\nabla_A X^0 ,\\
\nabla_A \bbeta^i & = \mathbbm E_A^M ((\BLambda\mathbbm R)^{-1}  \partial_M (\BLambda\mathbbm R))^{0i},\\
\nabla_A\bbtheta ^i & = \frac{1}{2} \mathbbm E_A^M \epsilon^{ijk} ((\BLambda\mathbbm R)^{-1} \partial_M (\BLambda \mathbbm R))^{jk},
\end{split} \end{equation}
such that ${\BLambda^\mu}_\nu = (e^{i \bbeta^i(\sigma) K_i} {)^\mu}_\nu$ and $\mathbbm R^{ij} = (e^{i\bbtheta^i(\sigma) J_i})^{ij}$.

We many now impose IH constraints $\nabla_0 \mathbbm X^i = 0$, which can be solved to give
\begin{equation}
\frac{\bbeta^i}{\bbeta}\tanh \bbeta = \frac{D_0 \mathbbm X^i}{D_0\mathbbm X^0},
\end{equation}
 thereby removing the boost Goldstones entirely. Notice that $\nabla_1 \mathbbm X^i$ have not been removed and may still appear in the action. 
We thus have the additional terms
\begin{equation}\begin{split}\label{spin spin} S_\text{rot} =\frac{1}{2i} \int d\tau  d\theta \mathbbm E \big( {I_{ij}} \nabla_0\bbtheta^i\nabla_1 \bbtheta^j + \vec w\cdot ( \nabla_0\vec \bbtheta \times \nabla_1\vec{ \mathbbm X}) \big),
 \end{split}\end{equation} 
 where $I_{ij}$ is a symmetric tensor and $\vec w$ a constant three-vector. The full spin-1/2 spinning-particle action is therefore given by a sum of~(\ref{massive higher s action}) and~(\ref{spin spin}). Just like the spin-0 case, $I_{ij}$ is the moment of inertia. Now, however, we have an additional term that has no spin-less counter-part, namely the terms involving $\vec w$. We interpret this  term as a kind of `spin-orbit' coupling; it couples the intrinsic spin of the object to the rigid-body rotation. Further, the form of $I_{ij}$ and $\vec w$ must respect the symmetry group $\cS\subset SO(3)$. Notice that only certain symmetry groups $\cS$ permit a nonzero $\vec w$. In particular $\vec w\neq 0$ is only possible if $\cS\subset U(1)$. By contrast, $I_{ij}\neq 0$ is always permitted.  


\section{Summary}

In this paper, we defined a generalized coset construction for systems with spontaneously broken Poincar\'e symmetry. The motivation was that when Poincar\'e symmetry is preserved, Goldstone's theorem is extremely restrictive: all Goldstones must be spin-0 bosonic particles with vanishing mass; however, when Poincar\'e symmetry is spontaneously broken, many more possibilities exist. In particular, Goldstone's theorem can be satisfied by excitations that possess any spin, may be bosonic or fermionic, and need not admit a particle or quasiparticle representation. Thus, systems with identical symmetry-breaking patterns may possess inequivalent Goldstone spectra. To illustrate this diversity, we chose to focus on the relativistic point particle. All relativistic point-particles have identical (or nearly identical) SSB patterns; yet they can have any mass $m\geq 0$ and have any spin $s\in \mathbb Z/2$. With our new-and-improved coset construction, we formulated effective actions for point-particles of arbitrary mass and spin. Along the way we identified a novel kind of inverse Higgs constraint that we termed the {\it on-shell IH constraint}, which arises when constructing EFTs for massless particles. This IH constraint bears a striking resemblance to the so-called dynamical IH constraint used to construct EFTs for fermi liquids~\cite{Ira, Ira 2}. Finally, we used this new coset construction to formulate EFTs for particles of arbitrary spin and finite spatial extent. In particular, we found a kind of spin-orbit coupling that describes interactions among physical rotation and quantum spin. 

With our new coset philosophy, the SSB pattern is no longer the only object of concern. Inspired by~\cite{Landry, Landry second sound}, we parameterized the full symmetry group with Goldstone fields defined on a worldvolume of our choosing and then imposed Gauge symmetries associated with the unbroken symmetry generators. Thus, the three ingredients that go into this novel coset construction are (1) identifying the SSB pattern, (2) choosing a worldvolume on which to construct the action and (3) picking a particular set of gauge symmetries. We were free to choose whatever gauge symmetries we liked; if we imposed the largest possible set of gauge symmetries, then the Goldstones associated with unbroken generators could be gauge-fixed to zero, thereby recovering the standard coset construction. To allow for a wide range of spins---including both bosonic and fermionic particles---we found that it was necessary to impose an $\cN=2s$ local SUSY on the worldline, where $s$ is the spin of the particle. This gauged SUSY is imposed at the level of the coset, meaning that once the gauge symmetries are specified, constructing an invariant action is just a matter of `turning the crank' and using the coset to read-off symmetry-invariant terms. Thus, this new coset construction provides the same advantages as the usual coset construction. 
 
We expect that this new coset construction will prove useful in a number of areas. In particular, the ordinary coset construction has proved to be a valuable tool when constructing EFTs for condensed matter systems~\cite{Zoology,More gapped Goldstones,coset}. Since our new coset construction allows for fermionic degrees of freedom, it is our hope and expectation that these new techniques will facilitate useful extensions of the coset construction to account for phases of matter with low-energy fermionic excitations. For example EFTs for fermi liquids, non-fermi liquids, and bad metals might now be realizable using this novel method of cosets. We also hope to extend this coset construction to allow for non-equilibrium EFTs defined on the Schwinger-Keldysh contour~\cite{Landry} and to allow for explicitly broken symmetries~\cite{Landry chemistry} that exhibit fermionic degrees of freedom. Finally, it would be of great interest to identify the rules that determine which gauge symmetries ought to be imposed. Since we are free to choose from an infinite set of gauge symmetries, it would be extremely useful to construct a dictionary between physical attributes of the system and emergent gauge symmetries in the coset construction.

\bigskip

\noindent{\bf Acknowledgments:} We would like to thank Alberto Nicolis and Lam Hui for their wonderful mentorship and Ioanna Kourkoulou for many insightful conversations. This work was partially supported by the US Department of Energy grant DE-SC0011941.

\appendix



\begin{thebibliography}{9}

\balance

\bibitem{Alberto Alberte}
L.~Alberte and A.~Nicolis,
``Spontaneously broken boosts and the Goldstone continuum,''
JHEP \textbf{07}, 076 (2020)
doi:10.1007/JHEP07(2020)076
[arXiv:2001.06024 [hep-th]].

  \bibitem{Weinberg}\label{Weinberg}
  Steven Weinberg,
  \textit{The quantum theory of fields. Vol. 2: Modern applications},
(Cambridge University Press, 1996).


\bibitem{Volkov}\label{Volkov}
 D.V. Volkov,
 ``{Phenomenological Lagrangians},''
Fiz.Elem.Chast.Atom.Yadra 4 (1973) 3-41.


 \bibitem{Ogievetsky}\label{Ogievetsky}
V. I. Ogievetsky,
 ``Nonlinear realizations of internal and space-time symmetries,''
in X-th winter school of theoretical physics in Karpacz, Poland. 1974.

 \bibitem{Ivanov and Ogievetsky}\label{Ivanov and Ogievetsky}
  E. Ivanov and V. Ogievetsky,
 ``The Inverse Higgs Phenomenon in Nonlinear Realizations,''
Teor.Mat.Fiz. 25 no. 2, (1975) 1050-1059.


\bibitem{Wheel} \label{Wheel}
  L.~V.~Delacr\'etaz, S.~Endlich, A.~Monin, R.~Penco and F.~Riva,
 ``(Re-)Inventing the Relativistic Wheel: Gravity, Cosets, and Spinning Objects,''
  JHEP {\bf 1411}, 008 (2014)
  doi:10.1007/JHEP11(2014)008
  [arXiv:1405.7384 [hep-th]].
  

\bibitem{WB1}
H.~Watanabe and T.~Brauner,
Phys. Rev. D \textbf{84}, 125013 (2011)
doi:10.1103/PhysRevD.84.125013
[arXiv:1109.6327 [hep-ph]].

\bibitem{WB2}
T.~Brauner and H.~Watanabe,
Phys. Rev. D \textbf{89}, no.8, 085004 (2014)
doi:10.1103/PhysRevD.89.085004
[arXiv:1401.5596 [hep-ph]].
    
    
    
\bibitem{Landry}
M.~J.~Landry,
``The coset construction for non-equilibrium systems,''
JHEP \textbf{07}, 200 (2020)
doi:10.1007/JHEP07(2020)200
[arXiv:1912.12301 [hep-th]].

\bibitem{Landry second sound}
M.~J.~Landry,
``Second sound and non-equilibrium effective field theory,''
[arXiv:2008.11725 [hep-th]].


\bibitem{Ira}
I.~Z.~Rothstein and P.~Shrivastava,
``Symmetry Realization via a Dynamical Inverse Higgs Mechanism,''
JHEP \textbf{05}, 014 (2018)
doi:10.1007/JHEP05(2018)014
[arXiv:1712.07795 [hep-th]].

\bibitem{Ira 2}
I.~Z.~Rothstein and P.~Shrivastava,
``Symmetry Obstruction to Fermi Liquid Behavior in the Unitary Limit,''
Phys. Rev. B \textbf{99}, no.3, 035101 (2019)
doi:10.1103/PhysRevB.99.035101
[arXiv:1712.07797 [cond-mat.str-el]].

\bibitem{vanHolten:1995qt}
J.~W.~van Holten,
``D = 1 supergravity and spinning particles,''
doi:10.1142/9789812830425-0010
[arXiv:hep-th/9510021 [hep-th]].

\bibitem{Landry quasicrystal}
M.~Baggioli and M.~Landry,
``Effective Field Theory for Quasicrystals and Phasons Dynamics,''
[arXiv:2008.05339 [hep-th]].


\bibitem{Zoology} 
  A.~Nicolis, R.~Penco, F.~Piazza and R.~Rattazzi,
  ``Zoology of condensed matter: Framids, ordinary stuff, extra-ordinary stuff,''
  JHEP {\bf 1506}, 155 (2015)
  doi:10.1007/JHEP06(2015)155
  [arXiv:1501.03845 [hep-th]].













\bibitem{Low} \label{Low}
  I.~Low and A.~V.~Manohar,
  ``Spontaneously broken space-time symmetries and Goldstone's theorem,''
  Phys.\ Rev.\ Lett.\  {\bf 88}, 101602 (2002)
  doi:10.1103/PhysRevLett.88.101602
  [hep-th/0110285].
  
  
    
\bibitem{More gapped Goldstones} 
  A.~Nicolis, R.~Penco, F.~Piazza and R.~A.~Rosen,
  ``{More on gapped Goldstones at finite density: More gapped Goldstones,}''
  JHEP {\bf 1311}, 055 (2013)
  doi:10.1007/JHEP11(2013)055
  [arXiv:1306.1240 [hep-th]].
  
  
\bibitem{UV completion without symmetry restoration} \label{UV completion without symmetry restoration} 
  S.~Endlich, A.~Nicolis and R.~Penco,
  ``{Ultraviolet completion without symmetry restoration,}''
  Phys.\ Rev.\ D {\bf 89}, no. 6, 065006 (2014)
  doi:10.1103/PhysRevD.89.065006
  [arXiv:1311.6491 [hep-th]].

 



\bibitem{coset} 
  A.~Nicolis, R.~Penco and R.~A.~Rosen,
  ``{Relativistic Fluids, Superfluids, Solids and Supersolids from a Coset Construction,}''
  Phys.\ Rev.\ D {\bf 89}, no. 4, 045002 (2014)
  doi:10.1103/PhysRevD.89.045002
  [arXiv:1307.0517 [hep-th]].

\bibitem{Joyce} \label{Joyce} 
  G.~Goon, A.~Joyce and M.~Trodden,
  ``{Spontaneously Broken Gauge Theories and the Coset Construction,}''
  Phys.\ Rev.\ D {\bf 90}, no. 2, 025022 (2014)
  doi:10.1103/PhysRevD.90.025022
  [arXiv:1405.5532 [hep-th]].


















   


\bibitem{Gates:1983nr}
S.~J.~Gates, M.~T.~Grisaru, M.~Rocek and W.~Siegel,
``Superspace Or One Thousand and One Lessons in Supersymmetry,''
Front. Phys. \textbf{58}, 1-548 (1983)
[arXiv:hep-th/0108200 [hep-th]].



\bibitem{vanHolten:1995ds}
J.~W.~van Holten,
``Propagators and path integrals,''
Nucl. Phys. B \textbf{457}, 375-407 (1995)
doi:10.1016/0550-3213(95)00520-X
[arXiv:hep-th/9508136 [hep-th]].

\bibitem{Frydryszak:1996mu}
A.~Frydryszak,
``Lagrangian models of the particles with spin: The First seventy years,''
doi:10.1142/9789812830425-0009
[arXiv:hep-th/9601020 [hep-th]].


\bibitem{Ikemori}
H. Ikemori,
``Superfield formulation of superparticles,''
Z. Phys. C \textbf{44}, 625-632 (1989)
doi:10.1007/BF01549083





\bibitem{Howe:1988ft}
P.~S.~Howe, S.~Penati, M.~Pernici and P.~K.~Townsend,
``Wave Equations for Arbitrary Spin From Quantization of the Extended Supersymmetric Spinning Particle,''
Phys. Lett. B \textbf{215}, 555-558 (1988)
doi:10.1016/0370-2693(88)91358-5



\bibitem{Edwards:2019eby}
J.~P.~Edwards and C.~Schubert,
``Quantum mechanical path integrals in the first quantised approach to quantum field theory,''
[arXiv:1912.10004 [hep-th]].

\bibitem{Marnelius:1993ba}
R.~Marnelius,
``Proper BRST quantization of relativistic particles,''
Nucl. Phys. B \textbf{418}, 353-378 (1994)
doi:10.1016/0550-3213(94)90251-8
[arXiv:hep-th/9309002 [hep-th]].




\bibitem{Landry chemistry}
M.~J.~Landry,
``Dynamical chemistry: non-equilibrium effective actions for reactive fluids,''
[arXiv:2006.13220 [hep-th]].









\end{thebibliography}
\end{document}